\documentclass[aps,prb,preprint,superscriptaddress]{revtex4-1}
\usepackage{braket}
\usepackage{graphicx}
\usepackage{hyperref}
\usepackage{float}

\begin{document}

\preprint{}

\title{
Origin and Evolution of Ultraflatbands in Twisted Bilayer Transition 
Metal Dichalcogenides: Realization of Triangular Quantum Dot Array
}


\author{Mit H. Naik}
\altaffiliation{Present address: Department of Physics, University of California at Berkeley, California 94720, USA and Materials Sciences Division, Lawrence Berkeley National Laboratory, Berkeley, California 94720, USA.}
\author{Sudipta Kundu}
\author{Indrajit Maity}
\author{Manish Jain}
\email{mjain@iisc.ac.in}
\affiliation{Center for Condensed Matter Theory, Department of Physics, Indian Institute of Science, Bangalore 560012, India}


\date{\today}

\begin{abstract}
Using a multiscale computational approach, we probe the origin and evolution of ultraflatbands in 
moir\'e superlattices of twisted bilayer MoS$_2$, a prototypical transition 
metal dichalcogenide. Unlike twisted bilayer graphene, we 
find no unique magic angles in twisted bilayer MoS$_2$ for flatband formation. 
Ultraflatbands form at the valence 
band edge for twist angles ($\theta$) close to 0$^\circ$ and 
 at \emph{both} the 
valence and conduction band edges for $\theta$ close to 60$^\circ$, and
have distinct origins. 
For $\theta$ close to 0$^\circ$, inhomogeneous hybridization
in the reconstructed moir\'e superlattice is sufficient to explain the formation of flatbands.
For $\theta$ close to 60$^\circ$, additionally, local strains cause 
the formation of modulating triangular potential wells such that electrons 
and holes are spatially separated.
This leads to multiple energy-separated ultraflatbands at the band edges closely resembling 
eigenfunctions of a quantum particle in an equilateral triangle well. 
Twisted bilayer transition metal dichalcogenides are thus suitable 
candidates for the realisation of ordered quantum dot array.



\end{abstract}


\maketitle

\section{Introduction}

Correlated insulating behaviour and unconventional superconductivity was
recently observed in twisted bilayer graphene (TBG) at a 'magic' angle of 1.1$^\circ$
\cite{Nature.Cao,Nature.Cao2,PNAS.MacD}. 
While the nature of superconductivity is still contested, 
formation of ultraflatbands near the Fermi level at this angle is essential to 
understanding these phenomena \cite{PRX.Po,PRL.AV,PRB.Choi,PRB.Su,
PRL.Gonz,PRL.Xu,PRL.Wu,PRB.Kennes}.  
Since this discovery, ultraflatbands
have been predicted in other 
twisted 2D materials \cite{PRB.Conte,PRB.Jeil,arxiv.Oleg,NL.Xian,PRB.Kang,arxiv.Kennes,arxiv.Zhao} including 
small angle twisted bilayer MoS$_2$ (TBM), a prototypical
transition metal dichalcogenide (TMD) \cite{PRL.Naik,PRL.MacD}.
For TBG, the bands flatten in a narrow range 
of 0.1$^\circ$ about 1.1$^\circ$ \cite{PRR.Carr, arxiv.Goodwin}, 
making their experimental realisation challenging. 
The existence or absence of similar unique 'magic' angles
in twisted TMDs has not been explored.
Ultraflatbands and localization also has significant 
implications on optical properties of the material \cite{Nature.Tran,Nature.Alexeev}. 

\begin{figure*}
\begin{center}
  \includegraphics[scale=0.18]{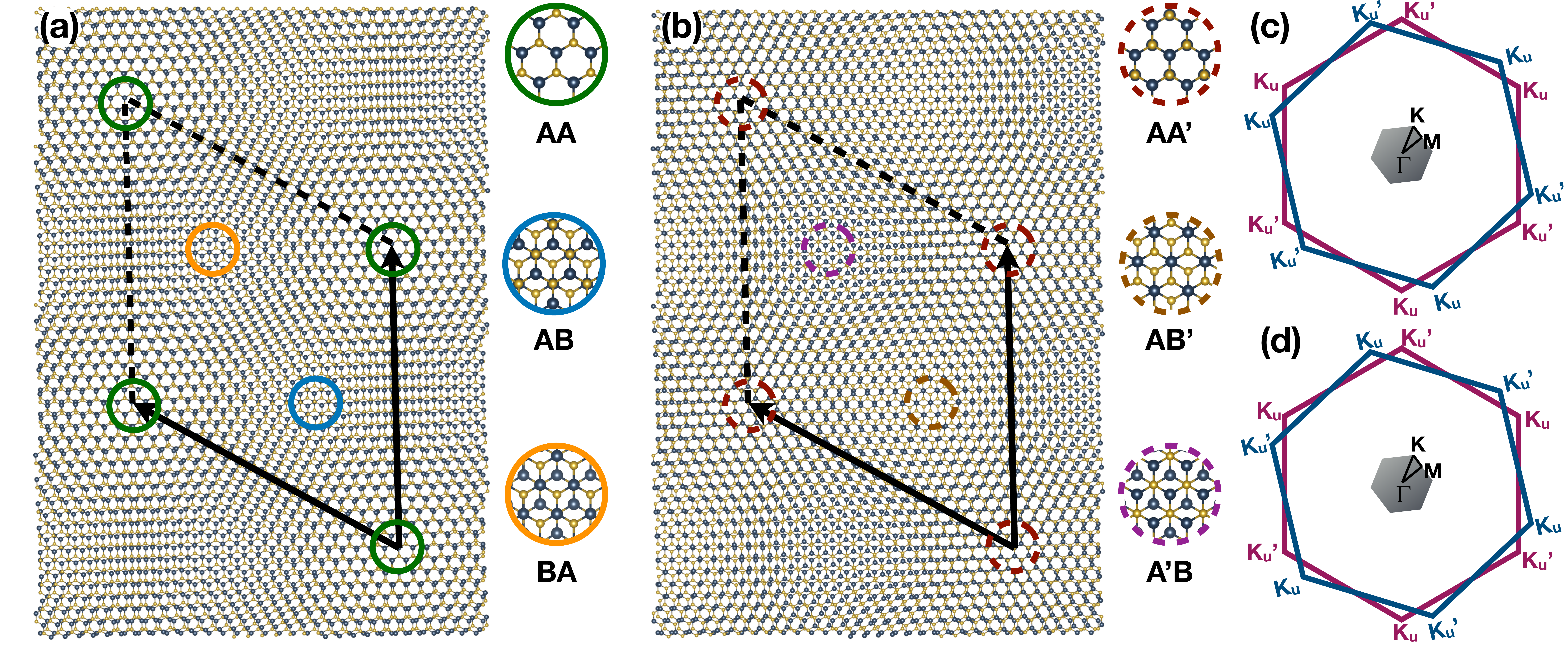}
\end{center}
\caption{\label{fig_struct}
(a) and (b) Structure of 2.65$^\circ$ and 57.35$^\circ$ rigidly twisted bilayer MoS$_2$. The
moir\'e pattern is composed of various stackings. The high-symmetry stackings
are identified using circles. The moir\'e superlattice vectors are shown with black arrows.
(c) and (d) Schematic of the unit-cell Brillouin zone (BZ) of the
bottom (in magenta) and top (blue) layer for a twist angle close to 0$^\circ$ and
60$^\circ$, respectively. The moir\'e BZ is shown in gray and the path connecting the
$\Gamma$, M and K points, along which the band structure is plotted, is marked.
}
\end{figure*}

\begin{figure}
\begin{center}
  \includegraphics[scale=0.24]{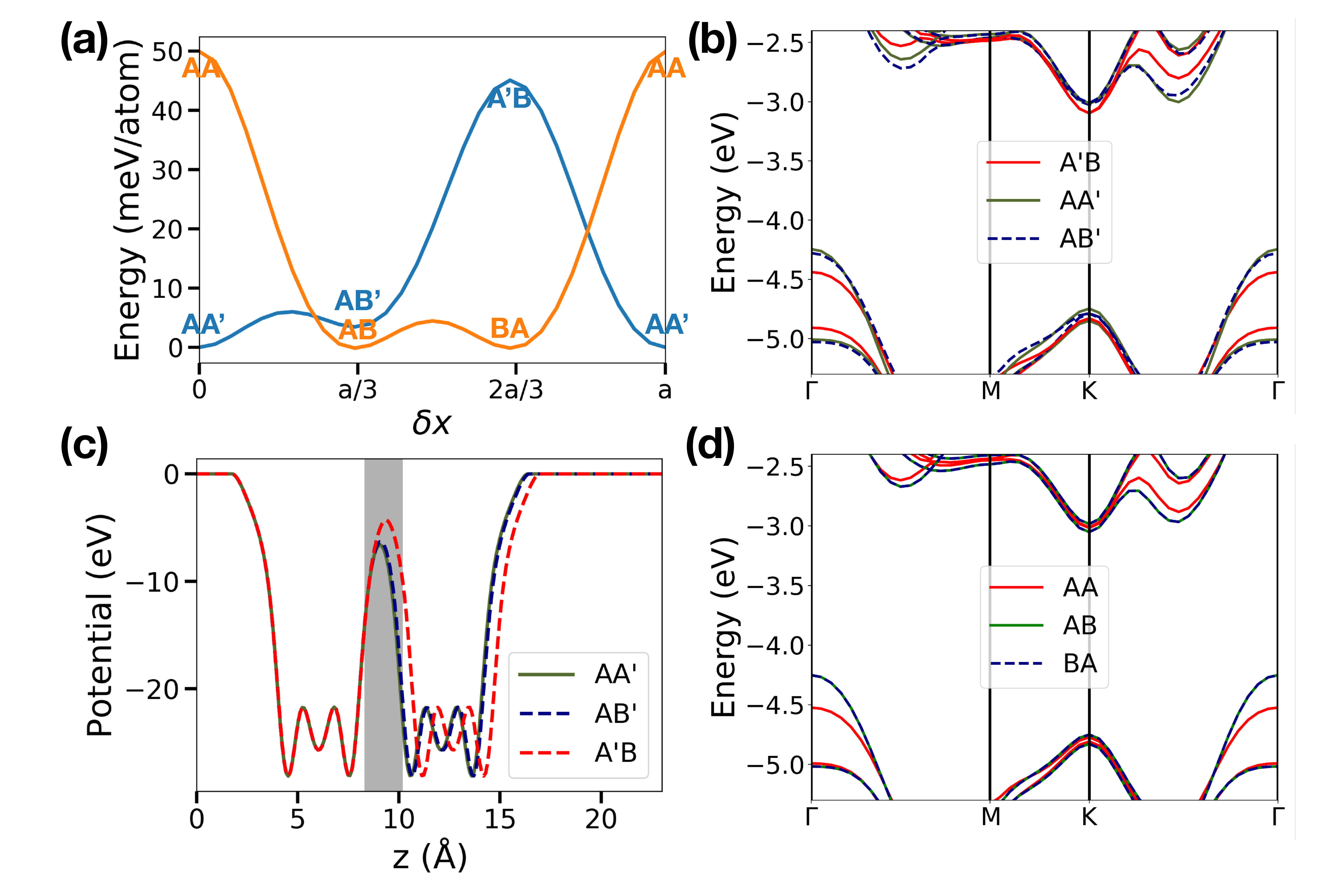}
\end{center}
\caption{\label{fig_uc}
(a) Relative energy of the stackings as a function of sliding the top layer with respect to the
bottom layer along the arm-chair direction in the unit-cell.
The starting configuration ($\delta x = 0$) is AA' and AA stacking for the blue and orange curve, respectively. 
(b) and (d) Electronic band structure of the isolated (unit-cell) high-symmetry stackings in Fig. 1.
(c) Planar-averaged DFT potential of the isolated high-symmetry stackings. The shaded area
marks the potential barrier between the layers.
In (b), (c) and (d) the spacing between the layers for each stacking is taken from the corresponding region
in the relaxed 2$^\circ$ and 58$^\circ$ twisted bilayer MoS$_2$ moir\'e superlattices.
}
\end{figure}

The structural properties of twisted TMDs
are remarkably different from TBG. Due to sublattice symmetry
breaking in the TMDs, distinct moir\'e patterns form for twist angles close to 0$^\circ$
and 60$^\circ$ \cite{PRL.Naik}. As the twist-angle approaches 0$^\circ$ or 60$^\circ$, the moir\'e
length scale increases. The sliding potential energy landscape
in twisted TMDs is more corrugated compared
to TBG leading to larger
deformation of the moir\'e superlattice  \cite{2D.Oleg,KC.Naik,arxiv.Maity}.
These deformations involve a change in the distribution of stackings and interlayer spacings from
the rigidly twisted structure \cite{PRL.Naik,arxiv.Maity,PRB.Carr_Soliton}.
The relaxed moir\'e pattern for twist angles ($\theta$)
close to 0$^\circ$ is similar to twisted bilayer graphene.
The low-energy bernal stacking occupies large equilateral triangle areas
separated by shear-strain solitons, while the
higher energy AA stacking region is reduced in size.
The relaxed moir\'e pattern for twist angle close to
60$^\circ$, on the other hand, is strikingly different \cite{arxiv.Maity,NN.Weston,AN.Rosenberger}.
For $\theta > 56^\circ$, the AA' stacking region occupies the largest area, transforming                
from an equilateral triangle to a Reuleaux triangle \cite{PRB.Carr_Soliton, arxiv.Maity}.                       
This leads to a reduction of the rotational symmetry                                                            
of the moir\'e superlattice from six-fold to three-fold. 

The electronic structure of twisted TMDs is strongly influenced by the structural                               
transformation of the moir\'e superlattice. Neglecting                                                          
structural relaxations in the simulation of these                                                               
systems leads to spurious localisation of flatbands \cite{PRL.Naik}.                                            
For $\theta = 3.5^\circ$, the flatband states close to the                                                      
valence band maximum localise to the bernal stacking sites, forming an extended                                 
hexagonal network in the moir\'e.                                                                               
For $\theta=56.5^\circ$, on the other hand, the valence band edge                                               
states have a smaller bandwidth and                                                                             
localise at the AA' stacking. However, the presence or absence of unique
'magic' angles for flatband formation                                                                           
and the influence of the releaux triangle pattern (for $\theta>56.5^\circ$) on the electronic                   
structure have not been explored and
are important to complete our understanding of twisted bilayer TMDs.                                            
First-principles study of the electronic structure of the relaxed moir\'e pattern for                           
$\theta > 56.5^\circ$ and $\theta < 3.5^\circ$ is computationally                                              
challenging due to the large number of atoms ($>1600$) involved in the simulation.

Quantum dots resemble artificial atoms with sizes in the order of nanometers.
\cite{Nat.Ashoori,Nat.Banin}
Quantum dots using 2D materials have several potential applications 
including quantum emission, design of solar cells and photocatalysis 
\cite{Science.Yu,AFM.Xu,AN.Deepesh,JAP.Gan,DT.Perumal,Small.Ha}. 
The current route to obtain quantum dots is through preparation of a colloidal
suspension of 2D material flakes \cite{JAP.Gan,AN.Deepesh,AFM.Xu,NJC.Lin}. 
This leads to poor control 
over the size and shape of quantum dots \cite{DT.Perumal}. 
Obtaining quantum dot array in a dry and systematic manner has been a 
challenge \cite{NJC.Lin}.
While the possibility of obtaining 
quantums dots in moir\'e patterns of twisted bilayers has been 
proposed \cite{Science.Yu,NL.Wu,arxiv.Shallcross}, 
explicit predictive 
calculations on the moir\'e pattern including crucial atomic relaxation effects are lacking. 
The scope of TBG for quantum dot applications is strongly                  
limited by the band gap and                                                                                    
formation of these flatbands only at 1.1$^\circ$ twist-angle \cite{PRR.Carr, arxiv.Goodwin}. 
Moir\'e patterns constructed using transition metal dichalcogenides (TMDs), on the other hand,                  
hold more promise with recent photoluminescence measurements showing signatures of localised
excitons in the moir\'e superlattice                                                                            
\cite{Nat.Seyler, Nat.Jin,Nat.Tran,Nature.Alexeev,arxiv.Regan,arxiv.Mauro}.




In this article, we use an efficient multiscale approach to study the origin 
and evolution of ultraflatbands in TBM. 
We establish, that unlike TBG, there are 
no unique magic angles in TBM for the formation of ultraflatbands.
Ultraflatbands form at the valence band edge 
for twist angles ($\theta$) close to $0^\circ$ and $60^\circ$. The electronic 
structure for $\theta$ close to $60^\circ$ is strikingly different from $0^\circ$. 
Multiple energy-separated ultraflatbands form at both the valence and conduction 
band edges for $\theta > 56^\circ$. 
Our calculations reveal a modulating
potential well in the moir\'e superlattice which leads to spatially separated electrons and holes.
The ordering, real-space distribution and the degeneracies of the ultraflatbands
at the valence band
edge are in excellent agreement with states of a quantum particle in an infinite equilateral triangle
potential well. The wavefunctions at the conduction band edge also resemble triangular well
states, with the degeneracies modified by valley degeneracies.
The ultraflatbands form due to two factors: 1) inhomogeneity in
the interlayer hybridization in the moir\'e superlattice due to variation in the interlayer spacing,
and 2) local strains due to soliton formation.
Furthermore, the local strains in each layer modify the electronic structure of
the optically active $K$ valleys of the unit-cell BZ. This could
result in confinement of excitons in the moir\'e pattern.

Twisted bilayer transition metal dichalcogenides are 
composed of distinct high-symmetry stackings for
twist angles close to 0$^\circ$ and 60$^\circ$. For twist angles close to
0$^\circ$, the high-symmetry stackings are AA,
AB (also referred to as B$^{\mathrm{Mo/S}}$ \cite{PRL.Naik})
and BA (B$^{\mathrm{S/Mo}}$). For twist angles close to
60$^\circ$, the moir\'e contains AA' (also referred to as AB), A'B (B$^\mathrm{S/S}$)
and AB' (B$^\mathrm{Mo/Mo}$)
high-symmetry stackings (Fig. \ref{fig_struct} (a) and (b)).
The formation of a moir\'e superlattice leads to shrinking of the unit-cell Brillouin zone (UBZ), 
as shown in Fig. \ref{fig_struct} (c) and (d).
The relative energy of the various stackings in the moir\'e
determines the relaxation pattern \cite{PRL.Naik,KC.Naik,PRB.Carr_Soliton, arxiv.Maity}.                        
The AA and A'B stackings are highest in energy (Fig. \ref{fig_uc} (a)) due to steric effects associated with 
S atoms of the bottom layer facing S atoms in the top layer. The AB and BA stackings are equal in energy, 
while the AA' stacking is lower in energy compared to AB'. This difference leads to different structural 
relaxations for twist angles close to 60$^\circ$ compared to those close to 0$^\circ$.
The interlayer spacing of the various stackings is also determined by the relative positions of the S atoms 
in the two layers. The AA and A'B stacking have a larger interlayer spacing, leading to a reduction in 
hybridisation between the layers. The valence band splittings at the $\Gamma$ point in the UBZ is controlled 
by the hybridisation between the layers \cite{PRB.Naik}. The different interlayer spacings for the different stackings leads to 
the variation in the unit-cell band structure shown in Fig. \ref{fig_uc} (b) and (d). 
The $\Gamma$ point valence band maximum (VBM) 
has the character of                                                                     
S-$p_z$ which make it more sensitive to interlayer spacing than the $K$ point VBM or 
conduction band minimum (CBM) (Mo-$d$ character)                     
\cite{PRB.Naik}.
The potential barrier between the two layers, as shown in Fig. \ref{fig_uc} (c), can be used a measure of the 
interlayer hybridisation \cite{PRB.Naik}. Larger interlayer spacing between the layers leads to an increase in the height of 
the barrier.

\section{Computational details}

The commensurate superlattices for twisted bilayer MoS$_2$ (TBM) are
constructed using the Twister code \cite{twister, PRL.Naik}.
The twist-angle, number of atoms in the moir\'e superlattice and length of the
superlattice vector in our simulations are provided in Table I.
The structural relaxations of TBM are performed with the LAMMPS \cite{LAMMPS,JCP.Plimpton} package
using intralayer Stillinger-Weber (SW) \cite{PRB.SW,Chap.SW}
and interlayer Kolmogorov-Crespi (KC) potential.
The force minimizations are performed using the conjugate gradient method with a tolerance of 10$^{-6}$  $\mathrm{eV/\AA}$.
The KC potential has been fit \cite{KC.Naik} to van der Waals (vdW) corrected DFT calculations.
The SW+KC forcefield relaxed
structure for TBM has been shown \cite{KC.Naik} to yield electronic structure in good agreement with
the vdW corrected DFT relaxed structure. 
The unit-cell
lattice constant of MoS$_2$ used in our calculations is 3.14 \AA.

The electronic structure calculations are performed on the 
relaxed moir\'e superlattice using density functional theory \cite{PR.Kohn} (DFT) calculations with the SIESTA 
\cite{SIESTA} package. 
The DFT wavefunctions are expanded in a double-$\zeta$ plus polarization basis.   
Norm-conserving pseudopotentials \cite{PRB.TM} and
the local density approximation to the exchange-correlation functional are 
employed. Van der Waals corrections only influence the interlayer spacing between the layers
in a bilayer sytem and do not influence the electronic band structure \cite{PRB.Naik}.
Since we are working with the relaxed structure of TBM, we do not use any vdW
correction while computing the electronic structure. We only sample the $\Gamma$
point in the moir\'e Brillouin zone (MBZ) to obtain the converged charge
density for the moir\'e superlattice calculations. For the
unit-cell simulations we use a $12\times12\times1$ sampling of the unit-cell BZ (UBZ).
A plane-wave energy cut-off of 250 Ry is used to generate the 3D grid for the simulation.
Spin-orbit coupling leads to a gap opening of $~$150 meV at the valence band edge at the $K$ point
in all unit-cell stackings.
Due to the large difference in the energy between the $\Gamma$ point VBM and the $K$ point VBM
in the UBZ, spin-orbit coupling effect does not influence the flatbands close to the valence
band edge in TBM \cite{PRL.Naik}. Hence, we do not include spin-orbit coupling in our electronic
structure calculations.

\begin{table*}[hbt!]
\centering
\begin{tabular}{c@{\hskip 0.07in}c@{\hskip 0.07in}c@{\hskip 0.07in}c}
\hline
\hline
 \multicolumn{1}{c@{\hskip 0.07in}}{Twist angle ($\theta$)}
& \multicolumn{1}{c@{\hskip 0.07in}}{Number of atoms}
& \multicolumn{1}{c@{\hskip 0.07in}}{Moir\'e superlattice size (\AA)}

\\
\hline
$1.54^\circ$ & 8322 & 117.1  \\
$2.0^\circ$ & 5514 & 95.3  \\
$2.65^\circ$ & 2814 & 68.1  \\
$2.88^\circ$ & 2382 & 62.6  \\
$57.12^\circ$ & 2382 & 62.6  \\
$57.35^\circ$ & 2814 & 68.1  \\
$58.0^\circ$ & 5514 & 95.3  \\
$58.46^\circ$ & 8322 & 117.1  \\
\hline
\hline
\end{tabular}
\caption{ Twist angles, number of atoms and sizes of the moir\'e superlattice vector in our
	calculations. The commensurate superlattices and constructed using the Twister code \cite{twister}.
}
\end{table*}


\section{Structural relaxations}

\begin{figure*}
\begin{center}
  \includegraphics[scale=0.34]{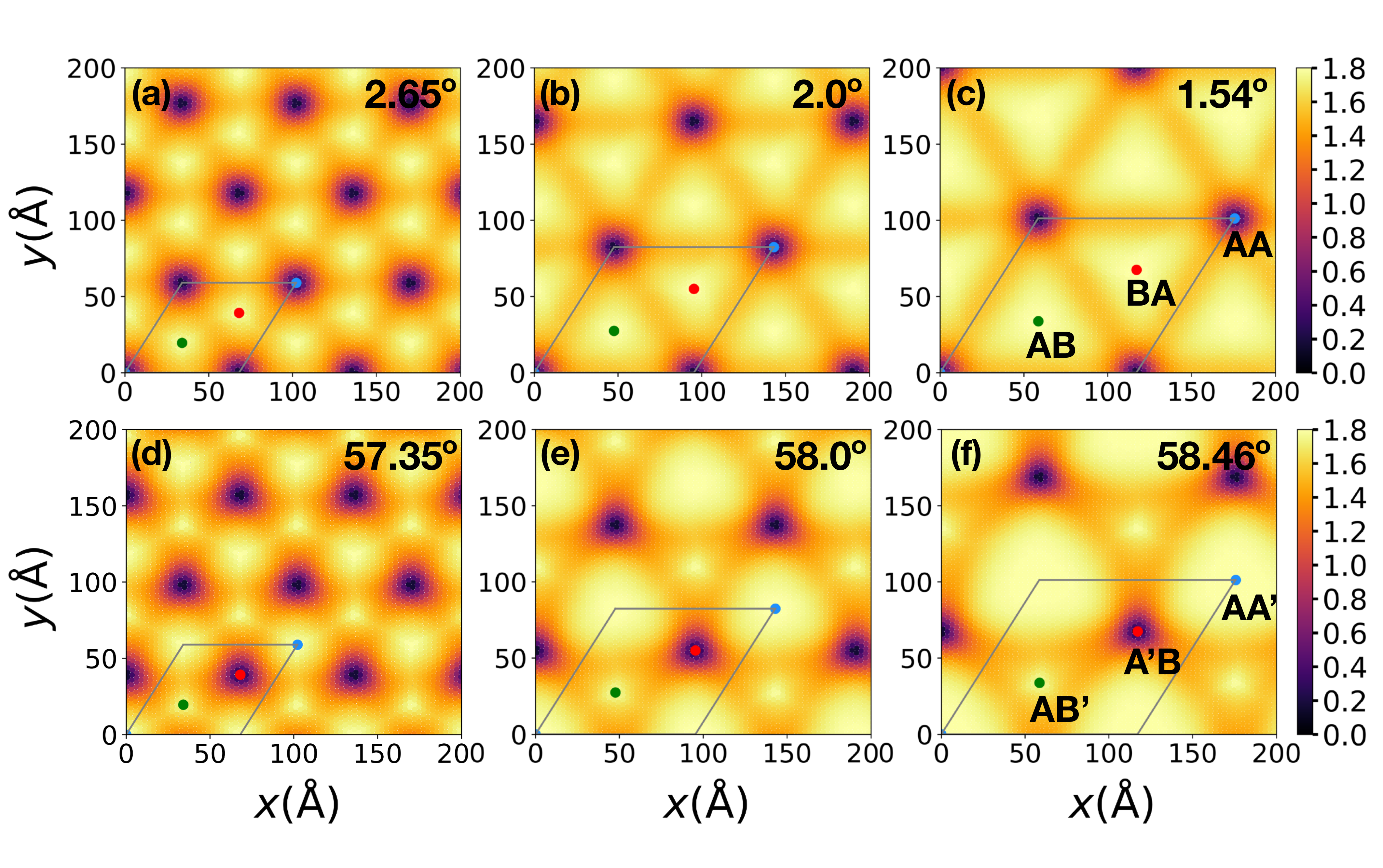}
\end{center}

 \caption{\label{fig_op}
	(a), (b) and (c) ((d), (e) and (f)) Order parameter distribution
	in 2.65$^\circ$, 2.0$^\circ$ and 1.54$^\circ$ (57.35$^\circ$, 58.0$^\circ$ and 58.46$^\circ$) 
	twisted bilayer MoS$_2$, respectively.
}
\end{figure*}

\begin{figure*}
\begin{center}
  \includegraphics[scale=0.34]{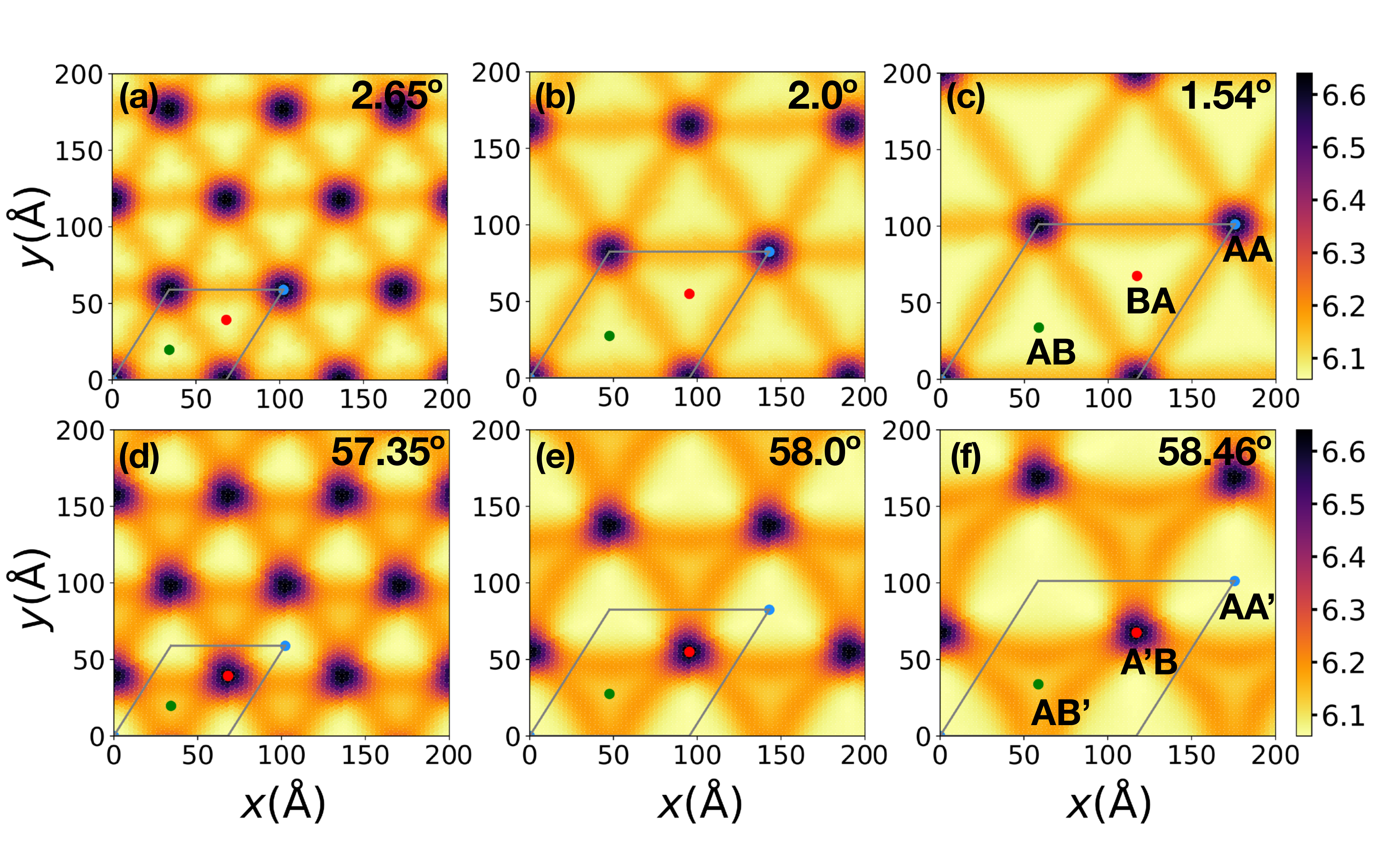}
\end{center}

 \caption{\label{fig_ils}
(a), (b) and (c) ((d), (e) and (f)) Interlayer spacing distribution
        in 2.65$^\circ$, 2.0$^\circ$ and 1.54$^\circ$ (57.35$^\circ$, 58.0$^\circ$ and 58.46$^\circ$)
        twisted bilayer MoS$_2$, respectively.
}
\end{figure*} 

\begin{figure}
  \centering
  \includegraphics[scale=0.43]{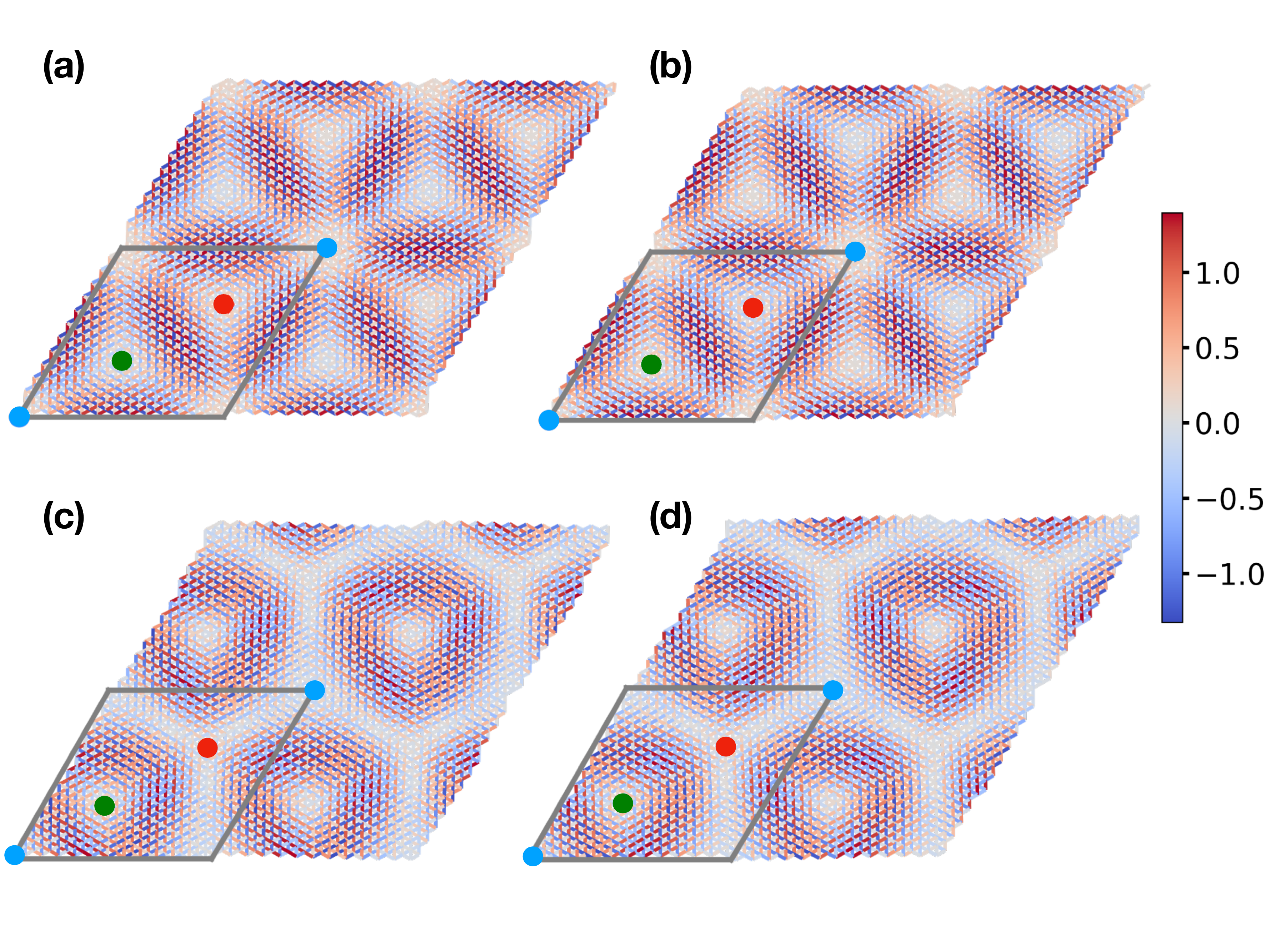}

 \caption{\label{strains}
  (a) and (b) ((c) and (d)) 
  Distribution of strains in the bottom and top layer of 2.65$^\circ$ (57.35$^\circ$) TBM.
  A line is drawn connecting each Mo to its six nearest neighbors. The color of the line shows the 
  strain in that direction. The dots in correspond to the location of
  high-symmetry stackings in the twisted bilayer. 
}
\end{figure}

Starting with the rigidly twisted structure, upon relaxation, 
the atoms in each layer locally shear in opposite directions                                   
to attain a lower energy stacking \cite{2D.Wijk,PRL.Naik}. This                                                         
shear leads to an in-plane strain in each layer                                                                 
of the moir\'e superlattice. The final                                                                          
relaxed pattern is hence a balance between the cost of in-plane strain and                                      
stacking energy gain.
To describe the redistribution of stackings upon relaxation we will use
order parameters \cite{2D.Oleg} (OP), $\vec{u}$ and $\vec{v}$ \cite{PRL.Naik}, to
describe the local stackings for twist angles close to 0$^\circ$ and 60$^\circ$, respectively.
$\vec{u}$ for a local stacking is defined as the displacement vector that transforms
the stacking to the highest energy AA stacking. 
$\vec{v}$ similarly transforms any local stacking in the
moir\'e to the A'B stacking. By definition, smaller the value of $|\vec{u}|$
or $|\vec{v}|$ less favourable the stacking. Fig. \ref{fig_op} shows the 
evolution of the $|\vec{u}|$ and  $|\vec{v}|$ as a function of twist-angle. 
The AB and BA regions remain triangles of equal area as                                                                           twist angle approaches 0$^\circ$. For twist angles approaching 60$^\circ$, on the other hand,                   
the AA' stacking region grows appreciably compared to other stackings.
The AA' stacking is lower in energy than AB'.                                                                       
Hence the area of AA' stacking in the moir\'e is larger and leads to a                                          
reduced three-fold symmetry around A'B. The domain wall                                                         
network has the shape of Reuleaux triangles \cite{PRB.Carr_Soliton, arxiv.Maity} in twist angles close          
to 60$^\circ$, as opposed to equilateral triangles in twist angles close                                
to 0$^\circ$. This contrast in the relaxation pattern leads to different electronic                             
structure for twist angles close to 0$^\circ$ and 60$^\circ$.
Furthermore, atomic displacements in the out-of-plane                                                           
direction lead to an undulating interlayer spacing in                                                           
the moir\'e \cite{arxiv.Maity, PRL.Naik}. 
The local interlayer spacing in the moir\'e (Fig. \ref{fig_ils}) 
is characteristic of the local stacking.                                                                        
The A'B stacking has the largest interlayer spacing due to Pauli repulsion 
owing to sulfur atoms in the bottom layer facing sulfur atoms in the top layer. 
The lattice reconstruction leads to the formation of an in inhomogeneous strain distribution in 
the moir\'e (Fig. \ref{strains}). The strain is localised along the soliton regions. Within 
each soliton region a network of tensile and compressive strain lines is formed. 
The magnitude and direction of these strains is switched between the top and bottom layer (Fig. \ref{strains}). 

\section{Electronic structure}

\subsection{Twist-angles close to 0$^\circ$}
\begin{figure}
  \includegraphics[scale=0.3]{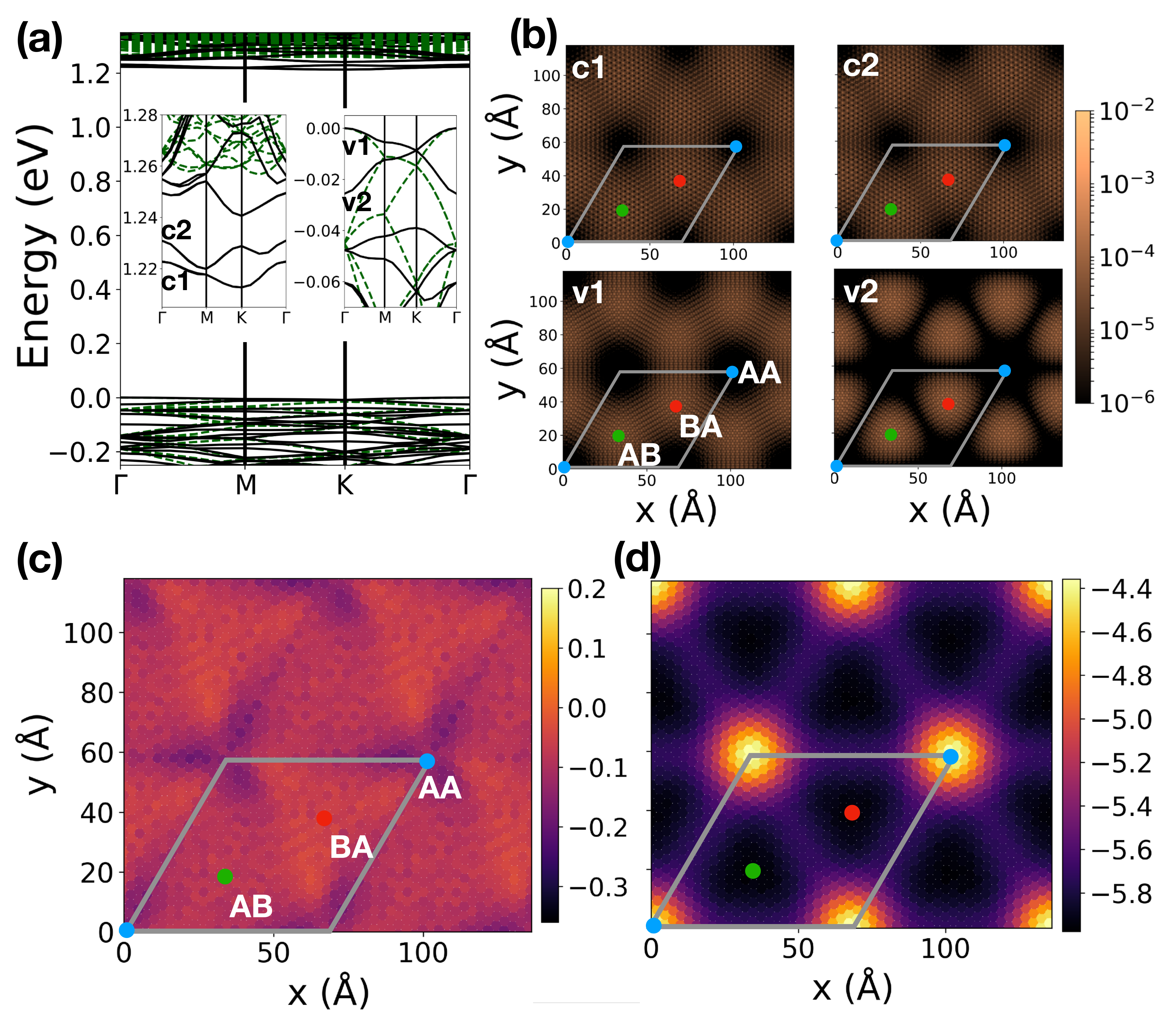}

 \caption{\label{2p65}
Electronic structure of 2.65$^\circ$ TBM. 
(a) Band structure (black) of 2.65$^\circ$ TBM.                                                     
Green dashed lines show band structure of                                                                       
purely AA' stacked bilayer MoS$_2$ in the same superlattice. The insets show                                    
plots of the valence and conduction band edges. (f) Charge density, $|\psi_{\Gamma}|^2$,                        
of the states labeled in the inset of (a). The charge density is averaged in                                    
the out-of-plane direction.                                                                                     
(c) $\Delta V(x_\mathrm{Mo},y_\mathrm{Mo})$ (in eV), for 2.65$^\circ$ TBM.                                     
(d) Distribution of the local potential barrier (in eV) between the layers, $V_\mathrm{barr}(x_\mathrm{Mo},y_\mathrm{Mo})$, in the moir\'e.                                                                                                           
 The extent of hybridization between the layers is inversely proportional                                       
 to the barrier.                                                                                                
}                                                                                                               
\end{figure}

\begin{figure}
 \begin{center}
  \includegraphics[scale=0.30]{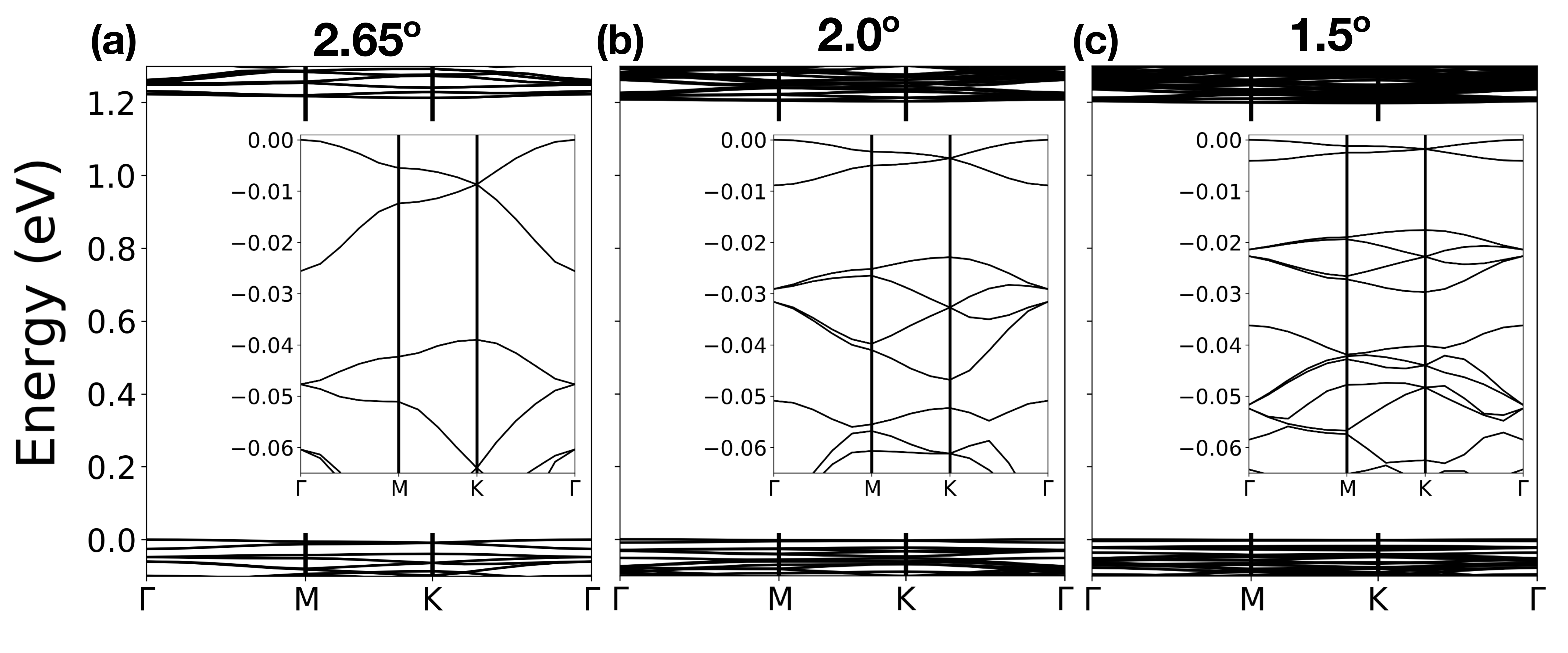}                                                                     
 \end{center}

 \caption{\label{bands_0}
 (a), (b) and (c) Band structure of 2.65$^\circ$, 2.0$^\circ$ and 1.5$^\circ$ TBM.
 The insets show enlarged plot of the bands close to the valence band edge.
}
\end{figure}

The electronic band structure plotted in the moir\'e Brillouine zone (MBZ) 
is a result of folding of bands from the UBZ of the individual layers (Fig. \ref{fig_struct} (c) and (d)).
It is hence essential to distinguish pure band-folding effects from that of flattening of the bands due to 
the moir\'e. We thus separately compute the band structure of pure AA' stacking in the 
same superlattice.
Fig. \ref{2p65} (a) shows bands of pure AA' stacking folded into the MBZ and also the band structure of the 
2.65$^\circ$ TBM. The bands of TBM
are clearly flatter than those of pure AA' stacking of same supercell size.
The bands flatten due to the localisation of the corresponding 
electronic states in real-space.  
The valence band edge states, v1 and v2 in inset of
Fig. \ref{2p65} (a) are degenerate at the $K$ point of the MBZ 
corresponding to the symmetry of the underlying lattice.
This degeneracy at the $K$ point is present in all twist angles close to 0$^\circ$ 
in our study (Fig. \ref{bands_0}). 
The dispersion of the v1 and v2 valence bands is similar to graphene due to the 
localisation of these states in a hexagonal pattern, avoiding the AA stacking region, 
as shown in Fig. \ref{2p65}.  
Similar to graphene, the degeneracy at the $K$ point is broken by the application of 
an external uniaxial strain to the moir\'e.   

\begin{figure}
 \begin{center}
  \includegraphics[scale=0.32]{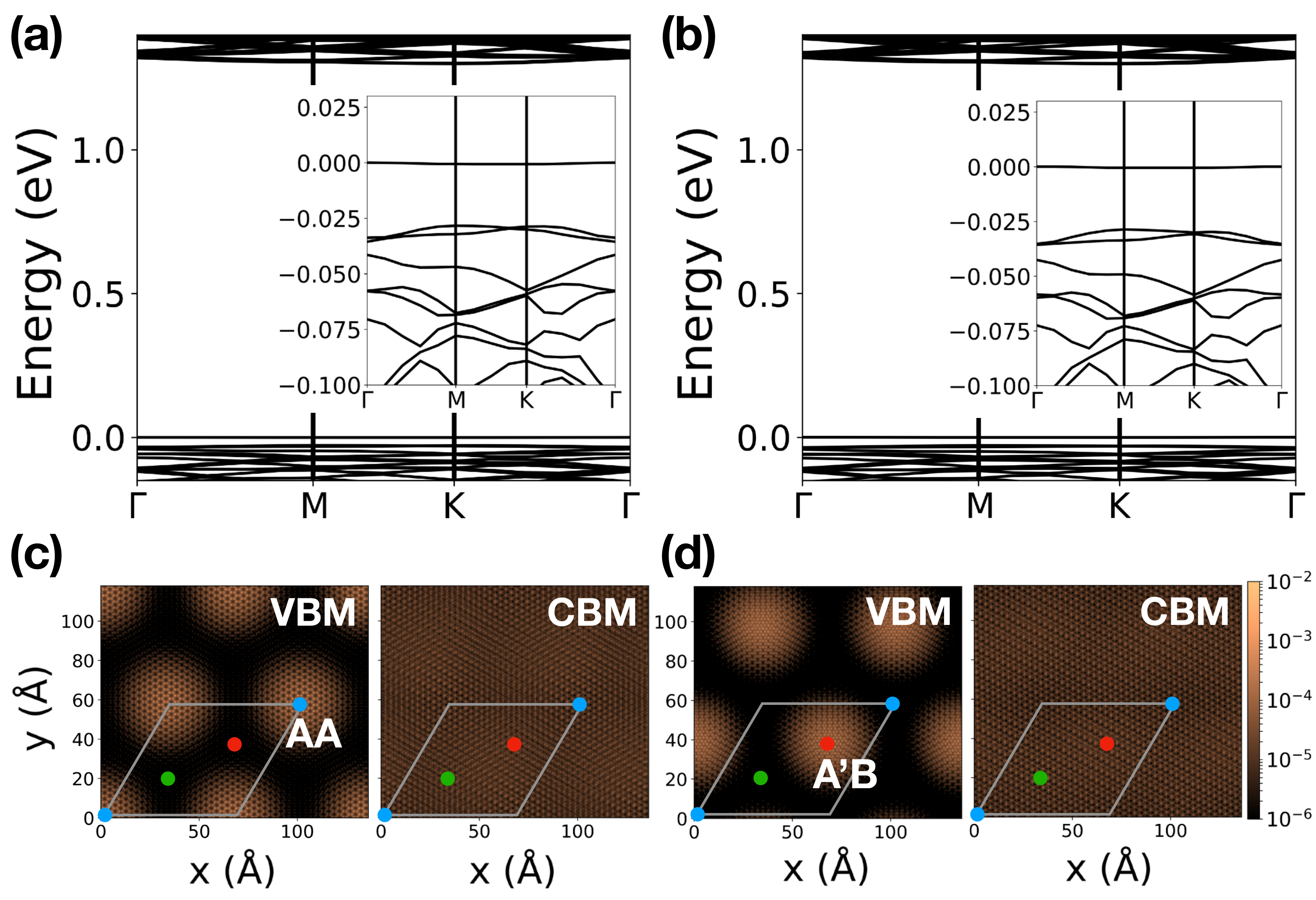}
 \end{center}

 \caption{\label{unrelaxed}
 Spurious localisation in rigidly-twisted structures. 
 (a) and (b) Band structure of 2.65$^\circ$ and 57.35$^\circ$ rigidly (unrelaxed) twisted bilayer MoS$_2$,                  
  respectively. The interlayer spacing is fixed at 6.3 \AA. The inset shows an enlarged           
  plot of the valence band edge.                                                                           
 (c) and (d) Distribution of valence band maximum and conduction band minimum 
 wavefunctions, $|\psi|^2$, of band structures shown in (a) and (b), respectively. 
}
\end{figure}

The localization of the bands at the valence band edge 
occurs due to inhomogeneous hybridization between the two layers 
in the moir\'e. 
As discussed previously \cite{PRL.Naik}, this can be 
qualitatively understood in terms of 
ordering of the VBM with respect to the vacuum level amongst different 
high-symmetry stackings (see Fig. \ref{fig_uc}). 
The ordering of the VBM among the stackings is determined by 
splittings at the $\Gamma$ point in the UBZ  (Fig. \ref{fig_uc}). 
Since the AA and A'B stackings have lower VBM (with respect to vacuum) 
compared to the other stackings, they cannot contribute to the VBM of the moir\'e superlattice. 
The conduction band edge lines up among the stackings 
 and provides no hint at a preferred localisation site. 
We can use the DFT potential barrier between the layers of a stacking as a measure of the 
interlayer hybridization for that stacking (Fig. \ref{fig_uc}). 
The extent of hybridization between the layers is 
inversely proportional to the DFT potential barrier between the layers.
To create a map of the local potential barrier between the layers of the moir\'e, we construct a 
Voronoi diagram using the Mo atoms of the bottom layer. The self-consistent DFT potential, 
$V(x,y,z)$, of the moir\'e 
is then planar averaged in each Voronoi cell individually to obtain $V(x_\mathrm{Mo},y_\mathrm{Mo}, z)$. 
($x_\mathrm{Mo}$, $y_\mathrm{Mo}$) are the coordinates of the Mo atoms in the bottom layer.  
The barrier potential is then obtained for each Voronoi cell giving $V_\mathrm{barr}(x_\mathrm{Mo},y_\mathrm{Mo})$.  
The distribution of $V_\mathrm{barr}(x_\mathrm{Mo},y_\mathrm{Mo})$ in Fig. \ref{2p65} (d) shows the inhomogeneous 
hybridisation between the layers. As expected, the barrier is lowest for AB and BA, and 
highest for the AA stacking region.
Neglecting structural relaxations leads to
a spurious energy-separated flatband at the valence band edge (Fig. \ref{unrelaxed}). 
This flatband is localised at the high-energy stacking regions, AA or
A'B, for twist angle close to 0$^\circ$ or 60$^\circ$,
respectively (Fig. \ref{unrelaxed}). The origin of this localisation is also 
inhomogeneous hybridisation in the rigidly-twisted moir\'e pattern. Due to the absence of 
varying interlayer spacing in the moir\'e, the ordering of the VBM among the 
unit-cell stackings is reversed \cite{PRL.Naik}. This leads to localisation at the AA stacking.

\begin{figure*}
  \centering
  \includegraphics[scale=0.29]{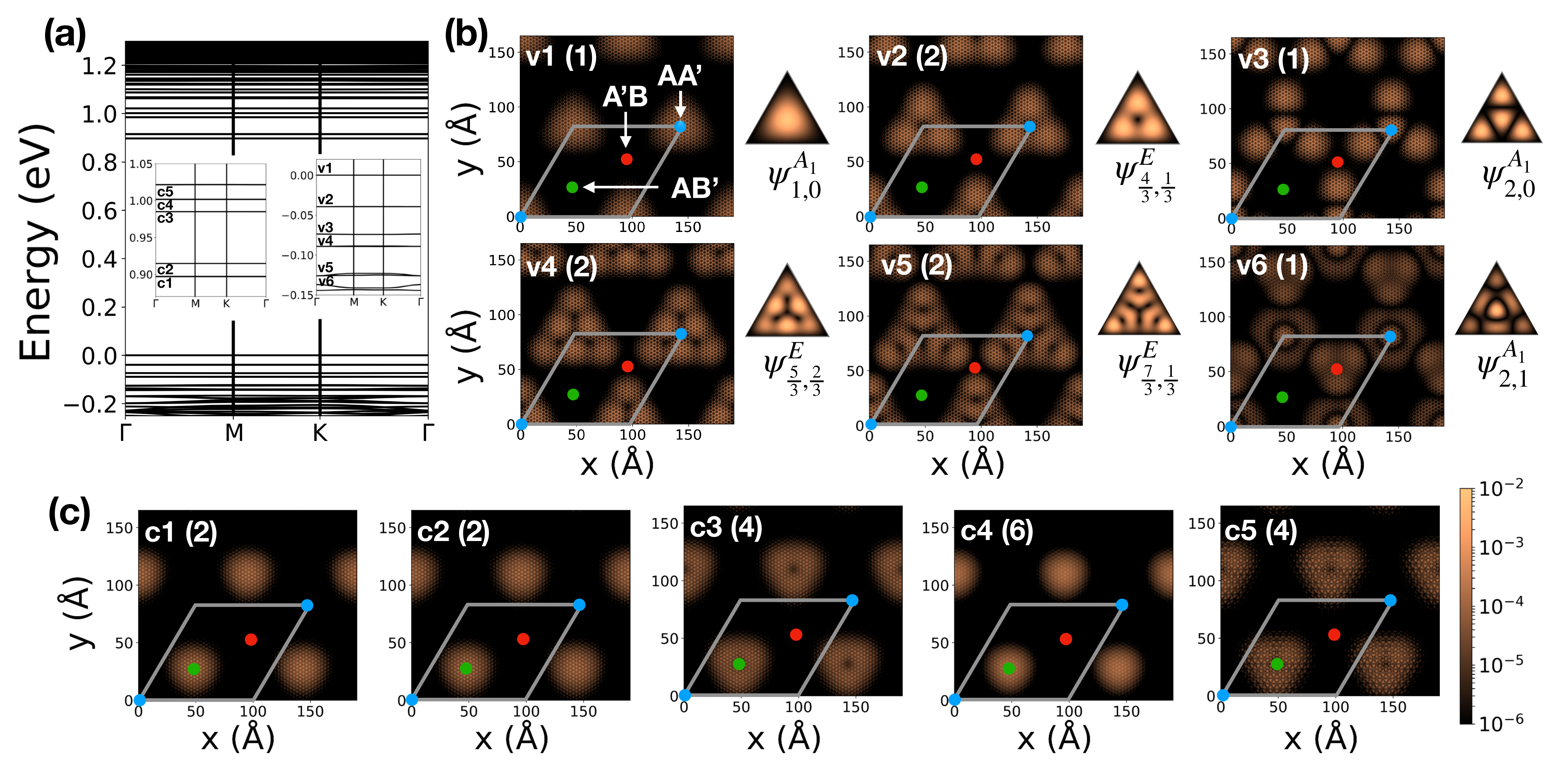}

 \caption{\label{ietpw}
 Electronic structure of 58$^\circ$ TBM.
 (a) Band structure of 58$^\circ$ TBM showing the multiple energy-separated
     ultraflatbands. The inset shows and labels the flatbands close to the valence and
     conduction band edges.
 (b) Distribution, $|\psi_{\Gamma}(\mathbf{r})^2|$, of the states labeled in
 (a), averaged in the out-of-plane ($z$) direction. The corresponding
  equilateral triangle quantum well wavefunctions
   of the ground state ($\psi^{A_1}_{1,0}$) and first five excited states  are shown alongside.
 (c) $|\psi_{\Gamma}(\mathbf{r})^2|$
     of the conduction states labeled in (a), c1-c5, averaged along the $z$-direction
     The degeneracies of the wavefunctions in (b) and (e) are shown in brackets.
}
\end{figure*}

\begin{figure}
  \centering
  \includegraphics[scale=0.27]{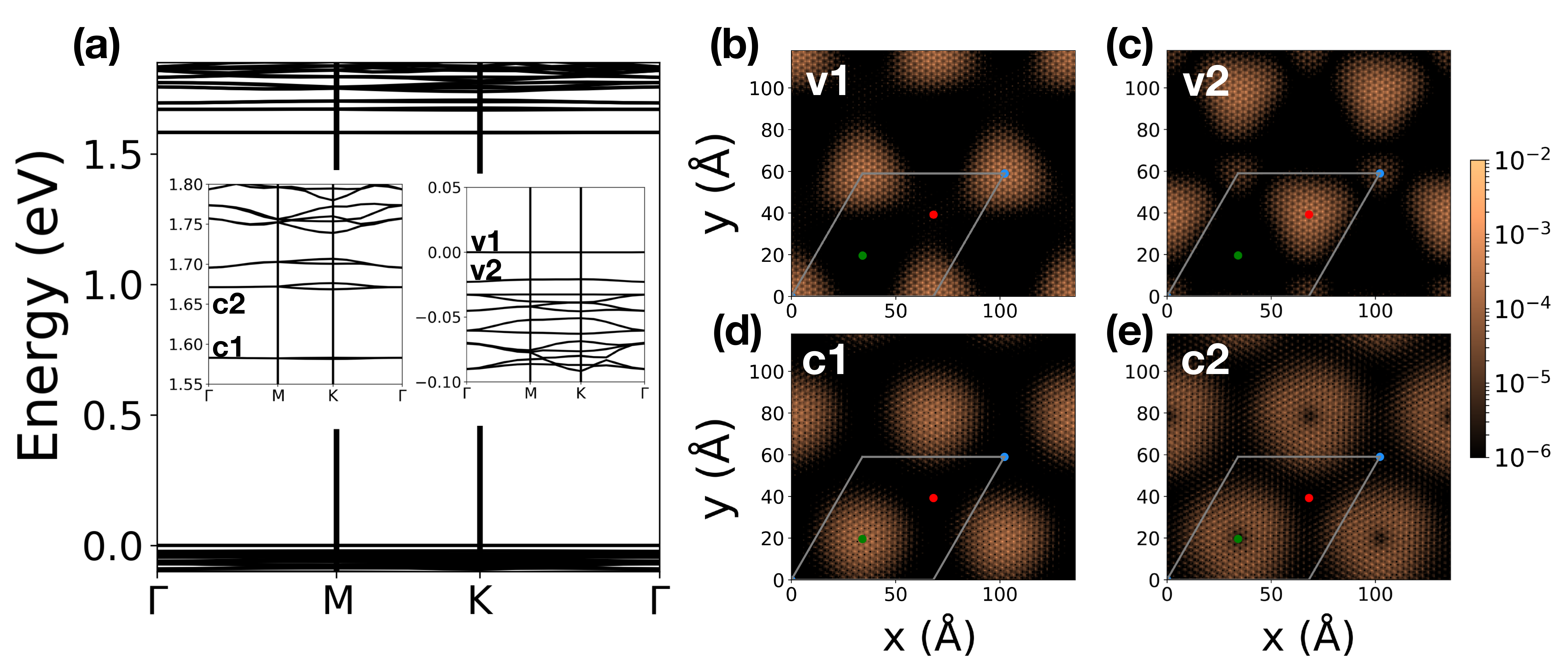}

 \caption{\label{es_mono}
  Electronic structure modification of individual layers. (a) Bandstructure of the bottom
        layer of relaxed 57.35$^\circ$ moir\'e superlattice. The inset shows the bands close to the
      valence and conduction band edge. The band edge states are folded in from the $K$ point of the 
unit-cell BZ.
  (b)  Distribution of valence band edge (v1 and v2) and
        conduction band edge (c1 and c2) wavefunctions corresponding
       to the states shown in (a).
}
\end{figure}

\begin{figure*}                                                                                                 
  \centering                                                                                                    
  \includegraphics[scale=0.34]{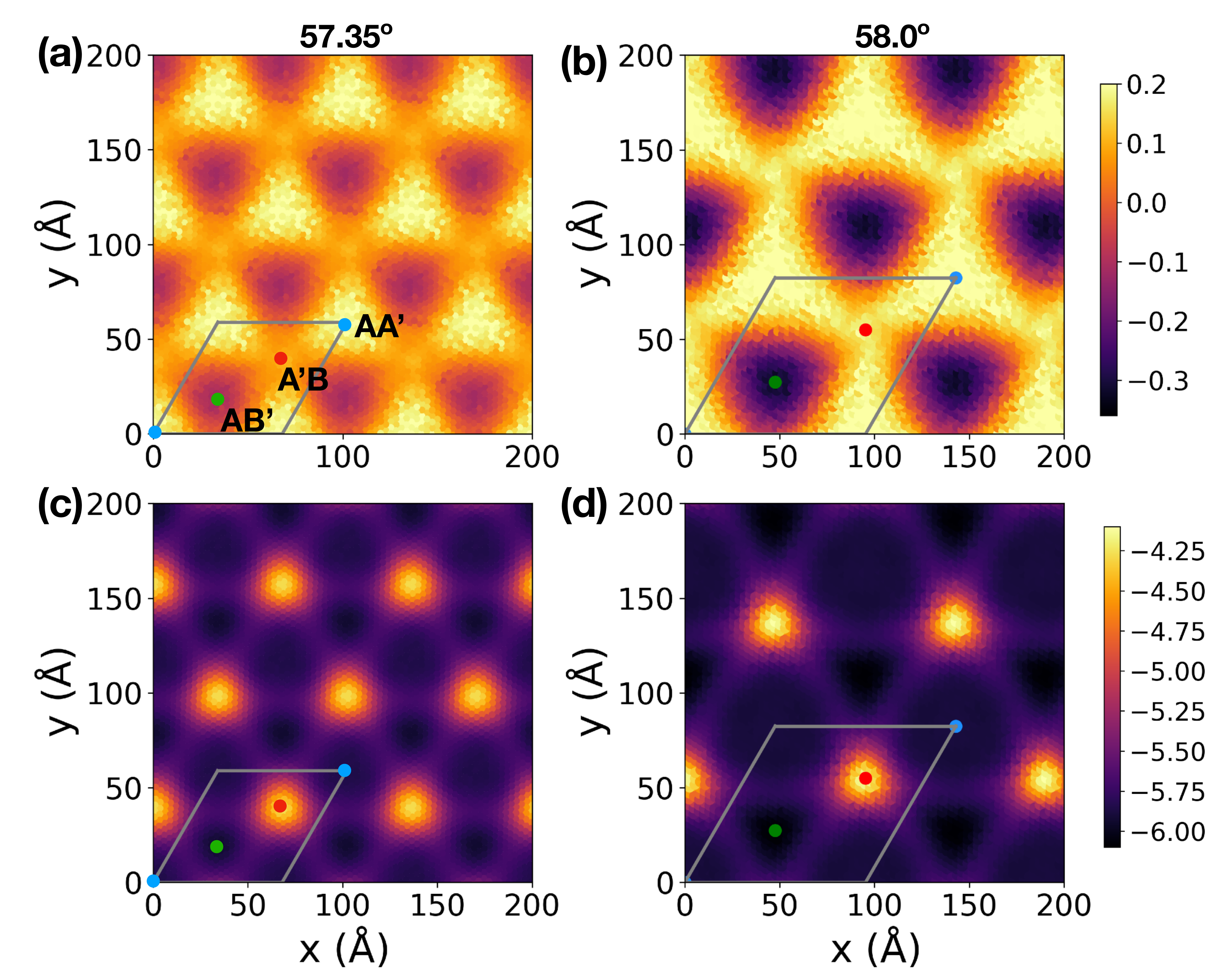}           
                                                                                                                
 \caption{\label{confining}                                                                                          
  (a) and (b) Confining potential, $\Delta V(x_\mathrm{Mo},y_\mathrm{Mo})$,                                     
     for 57.35$^\circ$ and 58$^\circ$ TBM, respectively.                                                        
  (c) and (d) Local potential barrier distribution (in eV) between the                                          
     layers, $V_\mathrm{barr}(x_\mathrm{Mo},y_\mathrm{Mo})$, in 57.35$^\circ$ and 58$^\circ$ TBM.               
}                                                                                                               
\end{figure*}

\subsection{Twist-angles close to 60$^\circ$}
In 58$^\circ$ relaxed TBM, several ultraflatbands form at the valence and conduction
band edges (Fig. \ref{ietpw} (a)) and are well separated in energy. The
states close to the valence band edge localise at the AA' stacking and conduction
band edge states at AB' (Fig. \ref{ietpw} (b) and (c)).
The spatial distribution of the wavefunctions
can be understood
in terms of an infinite equilateral triangle well potential.
A quantum particle in such a well (of side $a$) can be described using
two quantum numbers, $p$ and $q$. The energies are given
by $E_{p,q} = (p^2 + q^2 + pq)E_0$ \cite{JCE.Li}, $q$ takes the values
$0, 1/3, 2/3, 1, . . .$ and $p = q+1, q+2, q+3, . . .$. The ground
state, $E_{1,0} = E_0 =  2h^2/3ma^2$.
The eigenfunctions can be further labelled by $A_1$, $A_2$ and $E$. The
 $A_1$ and $A_2$ states are singly degenerate, while the $E$
states are doubly degenerate \cite{JCE.Li,JPA.HRK}.
In 58$^\circ$ TBM, the wavefunctions and degeneracies  
of the first six states (v1-v6) close to the valence band edge
are in excellent agreement with those of
the ground state and first five excited states of the infinite triangle
potential well (Fig. \ref{ietpw} (b)).
Similarly the conduction band edge states also agree well,
however the degeneracies do not follow those of the infinite triangle well.
c1 and c2 both are doubly degenerate and
correspond to $\psi^{A_1}_{1,0}$ of the triangular well.
These states have a valley degeneracy associated with them.
The four states that make up c1 and c2
are folded in from the $K$ and $K'$ points of the UBZ of the top and bottom layer.
The $K$ point wavefunction in the monolayer is strongly localised in the out-of-plane
direction \cite{PRB.Naik}.
Weak interlayer hybridization leads to a small gap opening
between the c1 and c2 states. c3 is four-fold degenerate,
and has an envelope function with a node at the center of the well
(Fig. \ref{ietpw} (c)) resembling $\psi^{E}_{\frac{4}{3},\frac{1}{3}}$. 
The six degenerate states are a result of the folding of the $\Lambda$ point 
(between $\Gamma$ and $K$) valley in the UBZ. The states close 
to the valence band edge fold only from the $\Gamma$ point of the UBZ. The 
degeneracies are hence unaffected and follow those of the infinite triangular well.          
It should be noted that in contrast to the ideal infinite triangle potential well, 
the potential in moir\'e is periodic and of finite depth. 
We thus expect only a few states close to the valence and conduction band edge to be 
confined in a triangular region.  

We also study the effect of the structural reconstruction on the electronic structure of the 
individual layers. Fig. \ref{es_mono} shows the electronic 
structure of the puckered 
bottom layer of relaxed 57.35$^\circ$ moir\'e. The strains (Fig. \ref{strains}) in the layer leads to 
localisation of the band edge states (Fig. \ref{es_mono}). As opposed to the bilayer, the valence and conduction bands 
in the monolayer fold in from the optically active $K$ point in the UBZ. The localisation of the $K$ point 
wavefunctions strongly suggests the modification of excitonic properties of these systems.
This could explain the recent discovery of moir\'e excitons in twisted bilayer 
TMDs \cite{Nat.Seyler, Nat.Jin,Nat.Tran,Nature.Alexeev,arxiv.Regan,arxiv.Mauro}.

\subsection{Origin of triangular quantum dots}
To understand the origin of the triangular quantum well potential, we plot the 
 distribution of $V_\mathrm{barr}(x_\mathrm{Mo},y_\mathrm{Mo})$
(Fig. \ref{confining} (c) and (d)), which shows the inhomogeneous hybridisation between the layers. 
As expected, the barrier is lowest for AA', AB' regions and 
highest for the A'B stacking region (Fig. \ref{fig_uc}). This suggests localisation of the 
valence band edge states at the AA' and AB' regions in the moir\'e. However, 
the valence band edge states localise at the AA' stacking                                                                  
alone and conduction band edge states at AB' (Fig. \ref{ietpw} (b) and (c)).
We find that in addition to the inhomogeneous hybridization,
a modulating potential is introduced in the moir\'e for twist angles close to
60$^\circ$ which explains the localisation pattern. 
To calculate the modulating potential, we first average the DFT potential 
 in a slab of length 17 \textrm{\AA} in the out-of-plane direction  
containing the bilayer, to obtain $V_\mathrm{M}(x,y)$.               
We then macroscopic average $V_\mathrm{M}(x,y)$, as discussed above,  
to obtain $V_\mathrm{M}(x_\mathrm{Mo},y_\mathrm{Mo})$. The confining   
potential with respect to AA' stacking is given by:                     
$\Delta V(x_\mathrm{Mo},y_\mathrm{Mo}) =                                 
V_\mathrm{M}(x_\mathrm{Mo},y_\mathrm{Mo}) - \bar{V}_\mathrm{AA'}$.        
Where $\bar{V}_\mathrm{AA'}$ is unit-cell averaged potential of AA' stacking.
$\Delta V(x_\mathrm{Mo},y_\mathrm{Mo})$ has a minimum 
at the AB' site, which confines the electrons (Fig. \ref{ietpw} (c)),
and a maximum at A'B and AA'. The inhomogeneous hybridization rules out localisation 
at A'B, hence inducing the holes to localise at the AA' stacking.   
The confining potential has the shape of an equilateral triangle (Fig. \ref{confining} (a) and (b)).
No modulation in $\Delta V(x_\mathrm{Mo},y_\mathrm{Mo})$ is
found for twist angles close to 0$^\circ$. 
These structures correspondingly have no localisation at the conduction band edge. 
Hence the electronic structure for twist angles close to 0$^\circ$ is very different
from that close to 60$^\circ$.

\begin{figure}
  \centering
  \includegraphics[scale=0.27]{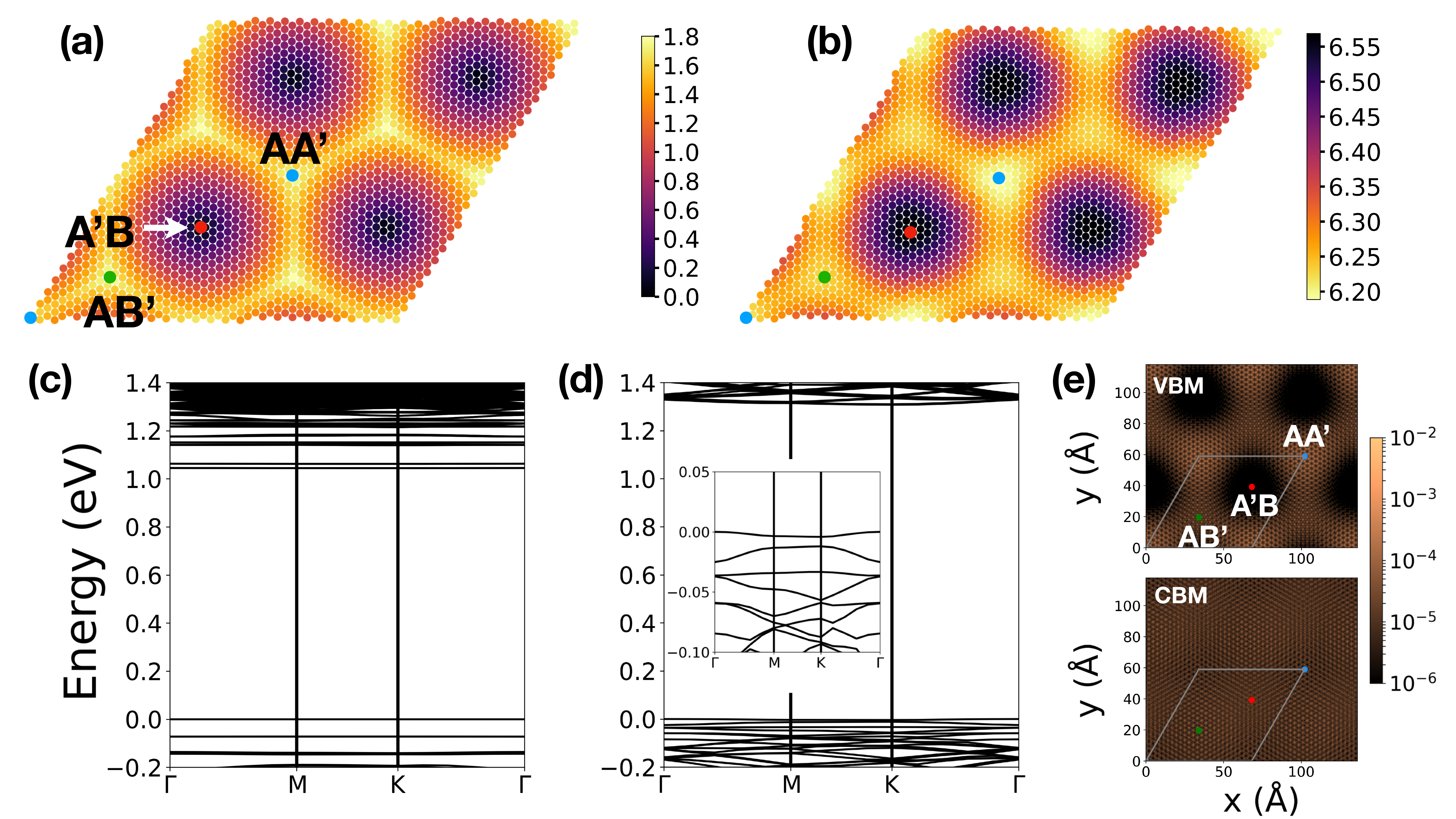}

 \caption{\label{constrained}
 Effect of constrained relaxation in 57.35$^\circ$ TBM.
 (a) and (b) Distribution of order-parameter and interlayer spacing in 57.35$^\circ$
 TBM with constrained relaxation (only out-of-plane relaxations are allowed), respectively.
 (c) and (d) Band structure of 57.35$^\circ$ TBM with full relaxation and
 constrained relaxation, respectively. 
 (e)  $|\psi|^2$ distribution of VBM and CBM wavefunctions, averaged along the out-of-plane direction, 
  for the band structure in (d).
}
\end{figure}

\begin{figure}
  \centering
  \includegraphics[scale=0.42]{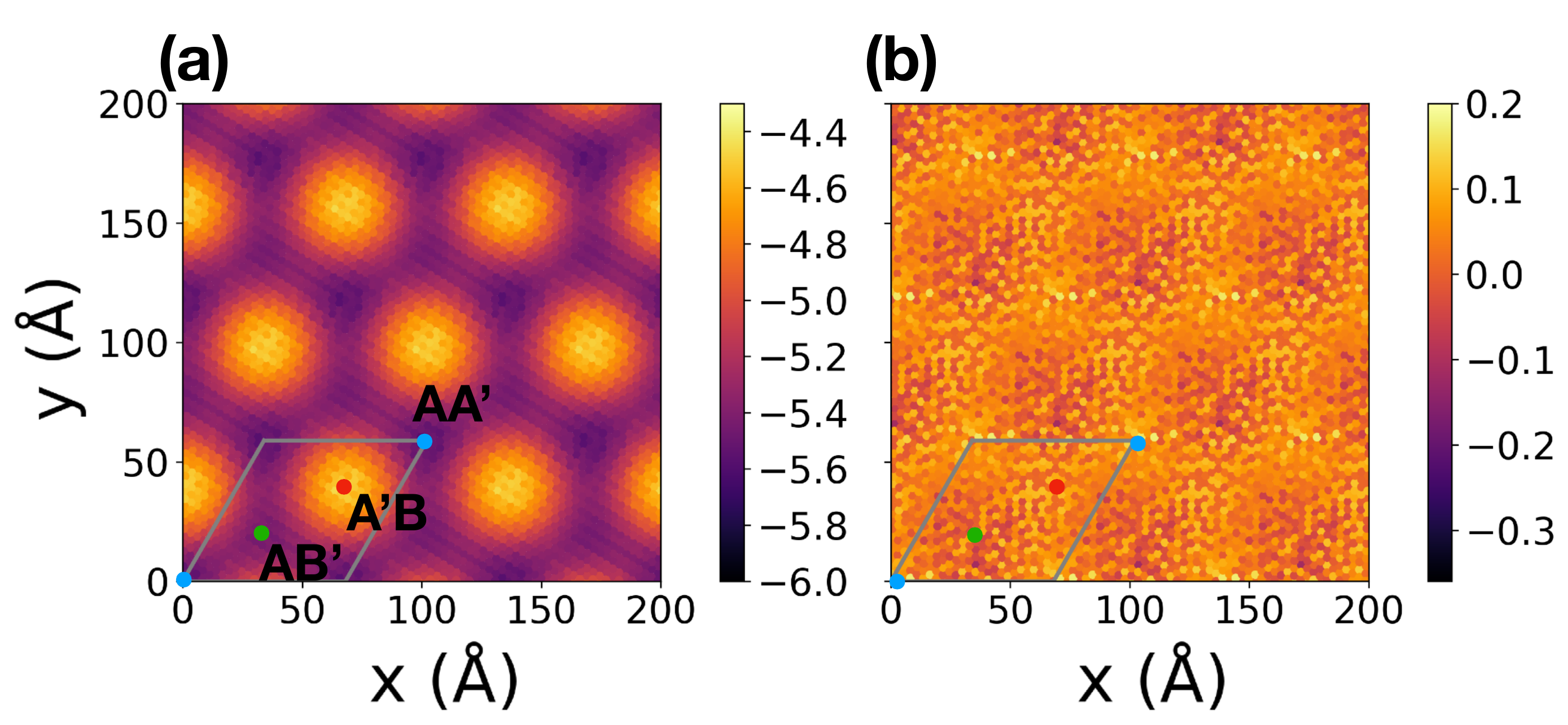}

 \caption{\label{constrainedV}
(a) and (b) Distribution of $V_\mathrm{barr}(x_\mathrm{Mo},y_\mathrm{Mo})$ 
 and $\Delta V(x_\mathrm{Mo},y_\mathrm{Mo})$
 in constrain relaxed (only out-of-plane relaxations are allowed) 57.35$^\circ$ TBM, respectively.
}                                                                                                               
\end{figure}

\begin{figure}
  \centering
  \includegraphics[scale=0.32]{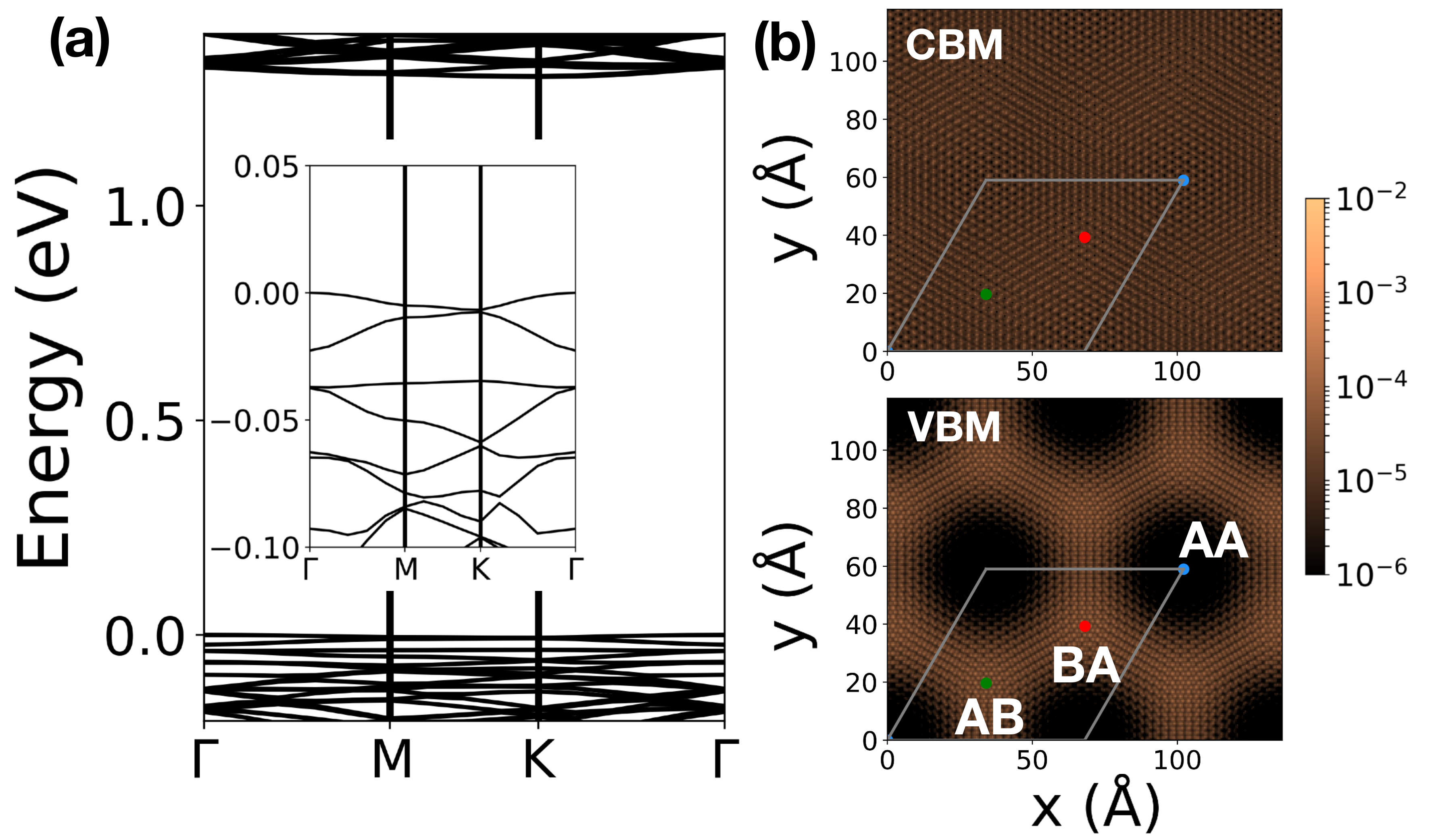}

 \caption{\label{constrainedAA}
 Effect of constrained relaxation in 2.65$^\circ$ TBM.
	(a) Band structure of 2.65$^\circ$ TBM with constrained relaxation of the 
 superlattice (only out-of-plane relaxations).
 (b)  $|\psi|^2$ distribution of VBM and CBM wavefunctions, averaged along the out-of-plane direction,
  for the band structure in (a).
}
\end{figure}

To understand the origin of the confining triangular potential in twist angles
close to 60$^\circ$ we probe the role of local strains in the moir\'e.
The strain is localised at the shear soliton regions in each layer (Fig. \ref{strains}).
We can construct a strain-free moir\'e by allowing atomic relaxations only in the
out-of-plane direction (from the rigidly twisted moir\'e superlattice). The interlayer spacing is allowed to
vary in this procedure (Fig. \ref{constrained} (b)),
hence the ordering of the VBM among the stackings is
preserved. We find that in this structure the multiple energy-separated
ultraflatbands in 57.35$^\circ$ vanish (Fig. \ref{constrained} (c) and (d)).
The conduction bands are delocalised due to the absence of a modulating
potential well and the valence bands are localised
at the AA' and AB' sites (Fig. \ref{constrained} (e)), as expected from the
hybridization arguments.  $V_\mathrm{barr}(x_\mathrm{Mo},y_\mathrm{Mo})$ and
$\Delta V(x_\mathrm{Mo},y_\mathrm{Mo})$ for this structure are shown in Fig. \ref{constrainedV}.
This clearly establishes in-plane relaxations, leading to strains, as the driving
mechanism for formation of the modulating potential well, which in turn
determines localisation of the wavefunctions. The localisation in 2.65$^\circ$ TBM, on the other hand,
is not affected by constrained relaxation (Fig. \ref{constrainedAA}).

\begin{figure*}
  \includegraphics[scale=0.5]{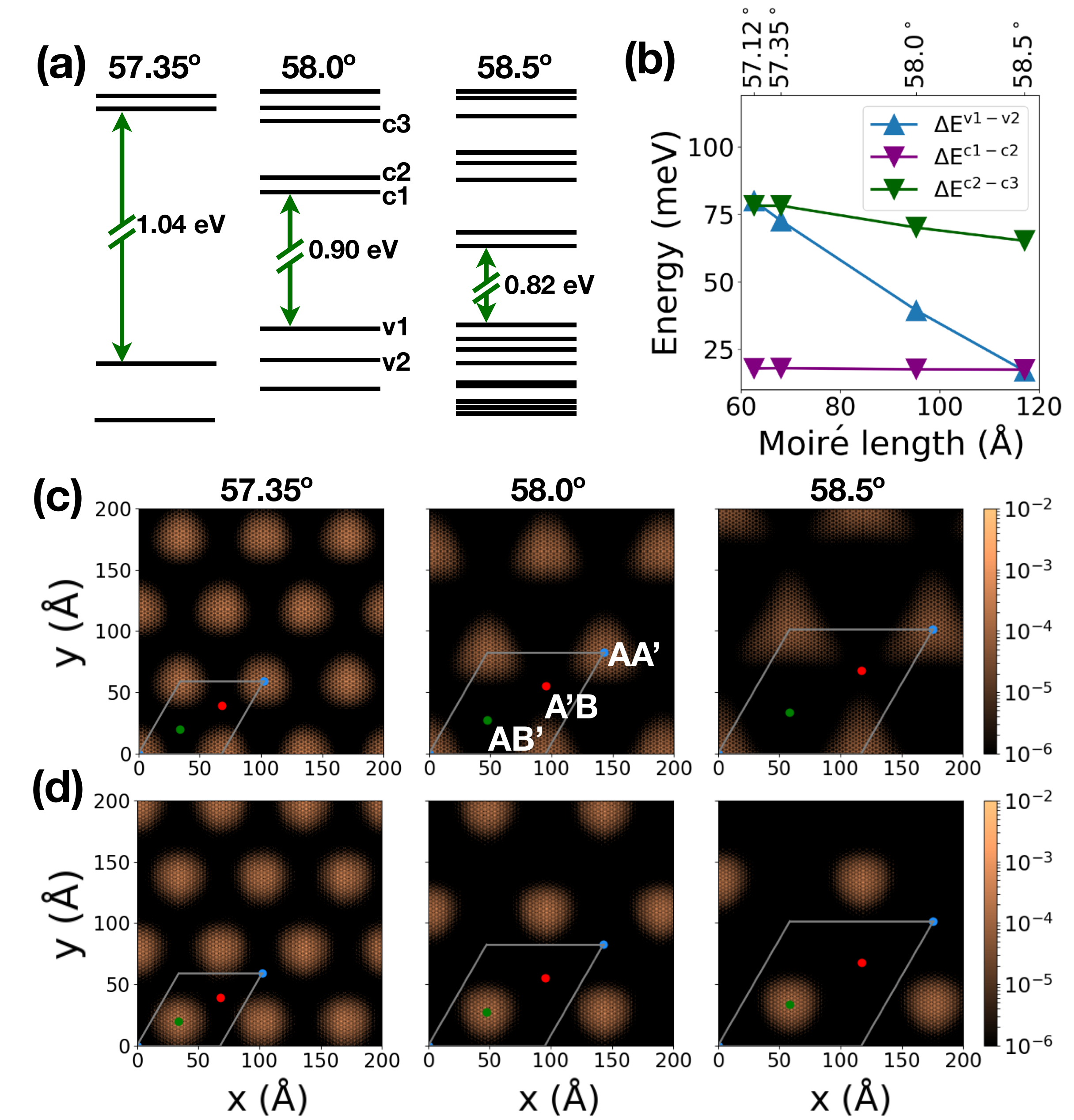}

 \caption{\label{tuning}
 Evolution of electronic structure with twist angle.
 (a) Evolution of ultraflatbands close to the valence and conduction band
     edge for twist angles 57.35$^\circ$, 58.0$^\circ$ and 58.5$^\circ$. The
     band gap is not to scale and is shown using a green arrow.
 (b) Evolution of the splittings between the first two states at the valence band edge,
     $\Delta \mathrm{E}^\mathrm{v1-v2}$, first ($\Delta \mathrm{E}^\mathrm{c1-c2}$)
     and second two states ($\Delta \mathrm{E}^\mathrm{c2-c3}$) at the conduction
     band edge.
  (c) ((d)) The wavefunction distribution of the v1 (c1) state in the moir\'e, averaged along the
      out-of-plane direction, for the three twist angles.
}
\end{figure*}

\begin{figure}
  \centering
  \includegraphics[scale=0.32]{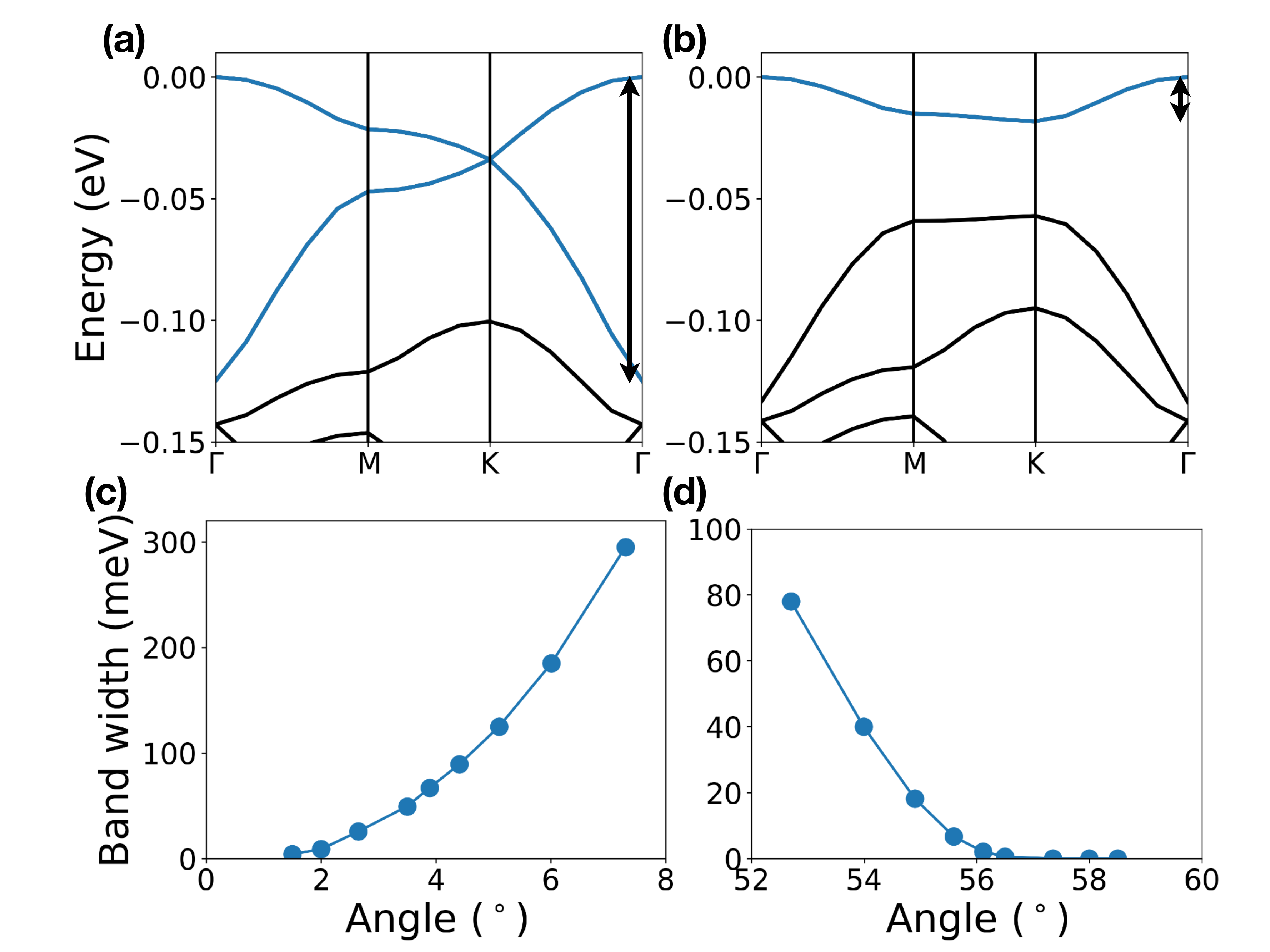}
 \caption{\label{bwidths}
(a) and (b) Bandstructure of 5.1$^\circ$ and 54.9$^\circ$ TBM, close to the
valence band edge, respectively. The bandwidth is measured for the first two bands in (a) and
the first band in (b).
 (c) and (d)  Band width (defined as shown in (a) and (b)) as a function of twist angle between MoS$_2$ layers.
}
\end{figure}

\begin{figure}                                                                                                  
  \centering                                                                                                    
  \includegraphics[scale=0.29]{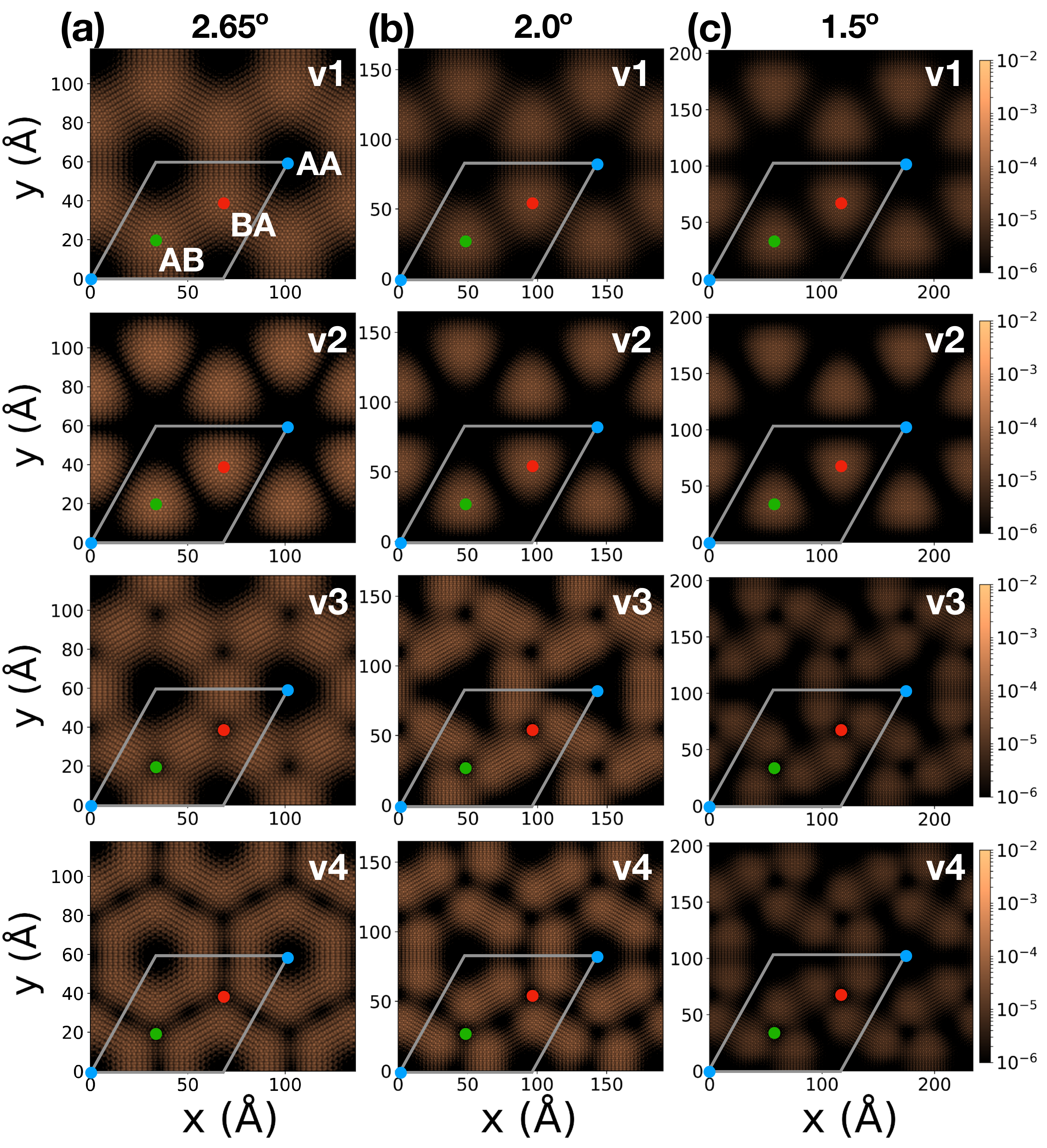} 
 \caption{\label{wfn0v}                                                                                       
(a), (b) and (c) Distribution of the valence band wavefunctions of TBM, averaged in the
 out-of-plane direction, for twist angles 2.65$^\circ$, 2.0$^\circ$ and
1.5$^\circ$, respectively.
 The $|\psi|^2$ distribution of the first four valence states are plotted at the $\Gamma$ point.
}                                                                                                               
\end{figure}

\subsection{Evolution of electronic structure with twist angle}

Fig. \ref{tuning} shows evolution of the flatbands for twist
angles 57.35$^\circ$, 58.0$^\circ$ and 58.5$^\circ$.
As the twist angle approaches 60$^\circ$, the AA' stacking area grows larger
than the other stackings (Fig. \ref{fig_op}).
As the area of confinement of holes increases, the
spacing between the flatbands close to valence band edge decreases as shown in
Fig. \ref{tuning} (b).
While the area of AB' region does not grow appreciably as the twist angle
approaches 60$^\circ$ (Fig. \ref{fig_op}), the confining potential depth
increases (Fig. \ref{confining}). The spacings between levels at the
conduction band edge are thus relatively unaffected (Fig. \ref{tuning} (b)).
Furthermore, the band gap of the moir\'e superlattice reduces as the twist 
angle approaches 60$^\circ$. While the band gap shown in Fig. \ref{tuning} (a) is the bandgap 
of the entire moir\'e, the band gap also varies locally in the moir\'e pattern.
The valence and conduction band edge energy for each local stacking is 
different due to the presence of the confining potential well. 
The valence band edge energy is highest at the AB stacking region and 
the conduction band edge is lowest at the AB' stacking. This spatial variation in 
the local density of states has been captured by scanning tunneling spectroscopy 
measurements of moir\'e superlattices \cite{SA.Zhang, NN.Zhang}.

The wavefunction localisation of the band edge states for 57.35$^\circ$, 58.0$^\circ$ and
58.5$^\circ$ twist angles is shown in Fig. \ref{tuning} (c) and (d).
The evolution of these states from 7.3$^\circ$ to 1.5$^\circ$ and 52.7$^\circ$ to 58.5$^\circ$
have been compiled into a movie available \cite{SM_ref} in the supplementary materials.
The states at the band edges can be regarded
as triangular quantum dot states for twist-angles greater than 56$^\circ$.
The real-space confinement of the wavefunctions leads to reduction in the band width of
these states. The band width of the first flatband at the valence band edge reduces monotonically
as the twist angle approaches 0$^\circ$ or 60$^\circ$ as shown in Fig. \ref{bwidths}.
The flatbands for twist angles greater than 56$^\circ$ are nearly dispersionless
with bandwidth less than 1 meV. This evolution clearly indicates the absence of
unique "magic" angles for flatband formation in TBM.

We also study the evolution of the first four flat bands at the valence band edge for
twist angles close to 0$^\circ$ (Fig. \ref{wfn0v}). The wavefunctions for twist angles
close to 0$^\circ$ always avoid the AA stacking region, as expected from the inhomogeneous hybridisation
in the moir\'e. The localisation patterns could be considered as solutions to a more complex
periodic potential well with minima at the AB and BA sites. The flatbands close to the
valence and conduction band edge for twist angle close to 60$^\circ$,
on the other hand, behave like triangular quantum dots as discussed above.

\section{Conclusion}

We have demonstrated the formation of
an array of triangular quantum dots in moir\'e patterns of TBM for
twist angles close to 60$^\circ$.
The holes and electrons are spatially separated which could
lead to long-lifetime confined excitons in this system.
By probing the origin and
evolution of the ultraflatbands we conclude that no special magic angles exist in
TBM like they do in twisted bilayer graphene.
This makes it easier to experimentally probe ultraflatbands in these systems.
The contrasting electronic
structure of twist angles close to 0$^\circ$ and 60$^\circ$
is due to an additonal modulating confining potential in twist angles close to
60$^\circ$. In-plane structural reconstruction of the moir\'e 
is responsible for the formation of this confining potential.
External strains can thus be used to engineer the confining potential
and flatbands in these systems. 

\begin{acknowledgments}
The authors thank the Supercomputer Education and Research Centre (SERC) at IISc
for providing the computational facilities.
\end{acknowledgments}




\begin{thebibliography}{61}%
\makeatletter
\providecommand \@ifxundefined [1]{%
 \@ifx{#1\undefined}
}%
\providecommand \@ifnum [1]{%
 \ifnum #1\expandafter \@firstoftwo
 \else \expandafter \@secondoftwo
 \fi
}%
\providecommand \@ifx [1]{%
 \ifx #1\expandafter \@firstoftwo
 \else \expandafter \@secondoftwo
 \fi
}%
\providecommand \natexlab [1]{#1}%
\providecommand \enquote  [1]{``#1''}%
\providecommand \bibnamefont  [1]{#1}%
\providecommand \bibfnamefont [1]{#1}%
\providecommand \citenamefont [1]{#1}%
\providecommand \href@noop [0]{\@secondoftwo}%
\providecommand \href [0]{\begingroup \@sanitize@url \@href}%
\providecommand \@href[1]{\@@startlink{#1}\@@href}%
\providecommand \@@href[1]{\endgroup#1\@@endlink}%
\providecommand \@sanitize@url [0]{\catcode `\\12\catcode `\$12\catcode
  `\&12\catcode `\#12\catcode `\^12\catcode `\_12\catcode `\%12\relax}%
\providecommand \@@startlink[1]{}%
\providecommand \@@endlink[0]{}%
\providecommand \url  [0]{\begingroup\@sanitize@url \@url }%
\providecommand \@url [1]{\endgroup\@href {#1}{\urlprefix }}%
\providecommand \urlprefix  [0]{URL }%
\providecommand \Eprint [0]{\href }%
\providecommand \doibase [0]{http://dx.doi.org/}%
\providecommand \selectlanguage [0]{\@gobble}%
\providecommand \bibinfo  [0]{\@secondoftwo}%
\providecommand \bibfield  [0]{\@secondoftwo}%
\providecommand \translation [1]{[#1]}%
\providecommand \BibitemOpen [0]{}%
\providecommand \bibitemStop [0]{}%
\providecommand \bibitemNoStop [0]{.\EOS\space}%
\providecommand \EOS [0]{\spacefactor3000\relax}%
\providecommand \BibitemShut  [1]{\csname bibitem#1\endcsname}%
\let\auto@bib@innerbib\@empty
\bibitem [{\citenamefont {Cao}\ \emph {et~al.}(2018{\natexlab{a}})\citenamefont
  {Cao}, \citenamefont {Fatemi}, \citenamefont {Fang}, \citenamefont
  {Watanabe}, \citenamefont {Taniguchi}, \citenamefont {Kaxiras},\ and\
  \citenamefont {Jarillo-Herrero}}]{Nature.Cao}%
  \BibitemOpen
  \bibfield  {author} {\bibinfo {author} {\bibfnamefont {Y.}~\bibnamefont
  {Cao}}, \bibinfo {author} {\bibfnamefont {V.}~\bibnamefont {Fatemi}},
  \bibinfo {author} {\bibfnamefont {S.}~\bibnamefont {Fang}}, \bibinfo {author}
  {\bibfnamefont {K.}~\bibnamefont {Watanabe}}, \bibinfo {author}
  {\bibfnamefont {T.}~\bibnamefont {Taniguchi}}, \bibinfo {author}
  {\bibfnamefont {E.}~\bibnamefont {Kaxiras}}, \ and\ \bibinfo {author}
  {\bibfnamefont {P.}~\bibnamefont {Jarillo-Herrero}},\ }\href@noop {}
  {\bibfield  {journal} {\bibinfo  {journal} {Nature}\ } (\bibinfo {year}
  {2018}{\natexlab{a}})}\BibitemShut {NoStop}%
\bibitem [{\citenamefont {Cao}\ \emph {et~al.}(2018{\natexlab{b}})\citenamefont
  {Cao}, \citenamefont {Fatemi}, \citenamefont {Demir}, \citenamefont {Fang},
  \citenamefont {Tomarken}, \citenamefont {Luo}, \citenamefont
  {Sanchez-Yamagishi}, \citenamefont {Watanabe}, \citenamefont {Taniguchi},
  \citenamefont {Kaxiras}, \citenamefont {Ashoori},\ and\ \citenamefont
  {Jarillo-Herrero}}]{Nature.Cao2}%
  \BibitemOpen
  \bibfield  {author} {\bibinfo {author} {\bibfnamefont {Y.}~\bibnamefont
  {Cao}}, \bibinfo {author} {\bibfnamefont {V.}~\bibnamefont {Fatemi}},
  \bibinfo {author} {\bibfnamefont {A.}~\bibnamefont {Demir}}, \bibinfo
  {author} {\bibfnamefont {S.}~\bibnamefont {Fang}}, \bibinfo {author}
  {\bibfnamefont {S.~L.}\ \bibnamefont {Tomarken}}, \bibinfo {author}
  {\bibfnamefont {J.~Y.}\ \bibnamefont {Luo}}, \bibinfo {author} {\bibfnamefont
  {J.~D.}\ \bibnamefont {Sanchez-Yamagishi}}, \bibinfo {author} {\bibfnamefont
  {K.}~\bibnamefont {Watanabe}}, \bibinfo {author} {\bibfnamefont
  {T.}~\bibnamefont {Taniguchi}}, \bibinfo {author} {\bibfnamefont
  {E.}~\bibnamefont {Kaxiras}}, \bibinfo {author} {\bibfnamefont {R.~C.}\
  \bibnamefont {Ashoori}}, \ and\ \bibinfo {author} {\bibfnamefont
  {P.}~\bibnamefont {Jarillo-Herrero}},\ }\href@noop {} {\bibfield  {journal}
  {\bibinfo  {journal} {Nature}\ } (\bibinfo {year}
  {2018}{\natexlab{b}})}\BibitemShut {NoStop}%
\bibitem [{\citenamefont {Bistritzer}\ and\ \citenamefont
  {MacDonald}(2011)}]{PNAS.MacD}%
  \BibitemOpen
  \bibfield  {author} {\bibinfo {author} {\bibfnamefont {R.}~\bibnamefont
  {Bistritzer}}\ and\ \bibinfo {author} {\bibfnamefont {A.~H.}\ \bibnamefont
  {MacDonald}},\ }\href@noop {} {\bibfield  {journal} {\bibinfo  {journal}
  {Proceedings of the National Academy of Sciences}\ }\textbf {\bibinfo
  {volume} {108}},\ \bibinfo {pages} {12233} (\bibinfo {year}
  {2011})}\BibitemShut {NoStop}%
\bibitem [{\citenamefont {Po}\ \emph {et~al.}(2018)\citenamefont {Po},
  \citenamefont {Zou}, \citenamefont {Vishwanath},\ and\ \citenamefont
  {Senthil}}]{PRX.Po}%
  \BibitemOpen
  \bibfield  {author} {\bibinfo {author} {\bibfnamefont {H.~C.}\ \bibnamefont
  {Po}}, \bibinfo {author} {\bibfnamefont {L.}~\bibnamefont {Zou}}, \bibinfo
  {author} {\bibfnamefont {A.}~\bibnamefont {Vishwanath}}, \ and\ \bibinfo
  {author} {\bibfnamefont {T.}~\bibnamefont {Senthil}},\ }\href {\doibase
  10.1103/PhysRevX.8.031089} {\bibfield  {journal} {\bibinfo  {journal} {Phys.
  Rev. X}\ }\textbf {\bibinfo {volume} {8}},\ \bibinfo {pages} {031089}
  (\bibinfo {year} {2018})}\BibitemShut {NoStop}%
\bibitem [{\citenamefont {Tarnopolsky}\ \emph {et~al.}(2019)\citenamefont
  {Tarnopolsky}, \citenamefont {Kruchkov},\ and\ \citenamefont
  {Vishwanath}}]{PRL.AV}%
  \BibitemOpen
  \bibfield  {author} {\bibinfo {author} {\bibfnamefont {G.}~\bibnamefont
  {Tarnopolsky}}, \bibinfo {author} {\bibfnamefont {A.~J.}\ \bibnamefont
  {Kruchkov}}, \ and\ \bibinfo {author} {\bibfnamefont {A.}~\bibnamefont
  {Vishwanath}},\ }\href {\doibase 10.1103/PhysRevLett.122.106405} {\bibfield
  {journal} {\bibinfo  {journal} {Phys. Rev. Lett.}\ }\textbf {\bibinfo
  {volume} {122}},\ \bibinfo {pages} {106405} (\bibinfo {year}
  {2019})}\BibitemShut {NoStop}%
\bibitem [{\citenamefont {Choi}\ and\ \citenamefont {Choi}(2018)}]{PRB.Choi}%
  \BibitemOpen
  \bibfield  {author} {\bibinfo {author} {\bibfnamefont {Y.~W.}\ \bibnamefont
  {Choi}}\ and\ \bibinfo {author} {\bibfnamefont {H.~J.}\ \bibnamefont
  {Choi}},\ }\href {\doibase 10.1103/PhysRevB.98.241412} {\bibfield  {journal}
  {\bibinfo  {journal} {Phys. Rev. B}\ }\textbf {\bibinfo {volume} {98}},\
  \bibinfo {pages} {241412} (\bibinfo {year} {2018})}\BibitemShut {NoStop}%
\bibitem [{\citenamefont {Su}\ and\ \citenamefont {Lin}(2018)}]{PRB.Su}%
  \BibitemOpen
  \bibfield  {author} {\bibinfo {author} {\bibfnamefont {Y.}~\bibnamefont
  {Su}}\ and\ \bibinfo {author} {\bibfnamefont {S.-Z.}\ \bibnamefont {Lin}},\
  }\href {\doibase 10.1103/PhysRevB.98.195101} {\bibfield  {journal} {\bibinfo
  {journal} {Phys. Rev. B}\ }\textbf {\bibinfo {volume} {98}},\ \bibinfo
  {pages} {195101} (\bibinfo {year} {2018})}\BibitemShut {NoStop}%
\bibitem [{\citenamefont {Gonz\'alez}\ and\ \citenamefont
  {Stauber}(2019)}]{PRL.Gonz}%
  \BibitemOpen
  \bibfield  {author} {\bibinfo {author} {\bibfnamefont {J.}~\bibnamefont
  {Gonz\'alez}}\ and\ \bibinfo {author} {\bibfnamefont {T.}~\bibnamefont
  {Stauber}},\ }\href {\doibase 10.1103/PhysRevLett.122.026801} {\bibfield
  {journal} {\bibinfo  {journal} {Phys. Rev. Lett.}\ }\textbf {\bibinfo
  {volume} {122}},\ \bibinfo {pages} {026801} (\bibinfo {year}
  {2019})}\BibitemShut {NoStop}%
\bibitem [{\citenamefont {Xu}\ and\ \citenamefont {Balents}(2018)}]{PRL.Xu}%
  \BibitemOpen
  \bibfield  {author} {\bibinfo {author} {\bibfnamefont {C.}~\bibnamefont
  {Xu}}\ and\ \bibinfo {author} {\bibfnamefont {L.}~\bibnamefont {Balents}},\
  }\href {\doibase 10.1103/PhysRevLett.121.087001} {\bibfield  {journal}
  {\bibinfo  {journal} {Phys. Rev. Lett.}\ }\textbf {\bibinfo {volume} {121}},\
  \bibinfo {pages} {087001} (\bibinfo {year} {2018})}\BibitemShut {NoStop}%
\bibitem [{\citenamefont {Wu}\ \emph {et~al.}(2018{\natexlab{a}})\citenamefont
  {Wu}, \citenamefont {MacDonald},\ and\ \citenamefont {Martin}}]{PRL.Wu}%
  \BibitemOpen
  \bibfield  {author} {\bibinfo {author} {\bibfnamefont {F.}~\bibnamefont
  {Wu}}, \bibinfo {author} {\bibfnamefont {A.~H.}\ \bibnamefont {MacDonald}}, \
  and\ \bibinfo {author} {\bibfnamefont {I.}~\bibnamefont {Martin}},\ }\href
  {\doibase 10.1103/PhysRevLett.121.257001} {\bibfield  {journal} {\bibinfo
  {journal} {Phys. Rev. Lett.}\ }\textbf {\bibinfo {volume} {121}},\ \bibinfo
  {pages} {257001} (\bibinfo {year} {2018}{\natexlab{a}})}\BibitemShut
  {NoStop}%
\bibitem [{\citenamefont {Kennes}\ \emph {et~al.}(2018)\citenamefont {Kennes},
  \citenamefont {Lischner},\ and\ \citenamefont {Karrasch}}]{PRB.Kennes}%
  \BibitemOpen
  \bibfield  {author} {\bibinfo {author} {\bibfnamefont {D.~M.}\ \bibnamefont
  {Kennes}}, \bibinfo {author} {\bibfnamefont {J.}~\bibnamefont {Lischner}}, \
  and\ \bibinfo {author} {\bibfnamefont {C.}~\bibnamefont {Karrasch}},\ }\href
  {\doibase 10.1103/PhysRevB.98.241407} {\bibfield  {journal} {\bibinfo
  {journal} {Phys. Rev. B}\ }\textbf {\bibinfo {volume} {98}},\ \bibinfo
  {pages} {241407} (\bibinfo {year} {2018})}\BibitemShut {NoStop}%
\bibitem [{\citenamefont {Conte}\ \emph {et~al.}(2019)\citenamefont {Conte},
  \citenamefont {Ninno},\ and\ \citenamefont {Cantele}}]{PRB.Conte}%
  \BibitemOpen
  \bibfield  {author} {\bibinfo {author} {\bibfnamefont {F.}~\bibnamefont
  {Conte}}, \bibinfo {author} {\bibfnamefont {D.}~\bibnamefont {Ninno}}, \ and\
  \bibinfo {author} {\bibfnamefont {G.}~\bibnamefont {Cantele}},\ }\href
  {\doibase 10.1103/PhysRevB.99.155429} {\bibfield  {journal} {\bibinfo
  {journal} {Phys. Rev. B}\ }\textbf {\bibinfo {volume} {99}},\ \bibinfo
  {pages} {155429} (\bibinfo {year} {2019})}\BibitemShut {NoStop}%
\bibitem [{\citenamefont {Chebrolu}\ \emph {et~al.}(2019)\citenamefont
  {Chebrolu}, \citenamefont {Chittari},\ and\ \citenamefont {Jung}}]{PRB.Jeil}%
  \BibitemOpen
  \bibfield  {author} {\bibinfo {author} {\bibfnamefont {N.~R.}\ \bibnamefont
  {Chebrolu}}, \bibinfo {author} {\bibfnamefont {B.~L.}\ \bibnamefont
  {Chittari}}, \ and\ \bibinfo {author} {\bibfnamefont {J.}~\bibnamefont
  {Jung}},\ }\href {\doibase 10.1103/PhysRevB.99.235417} {\bibfield  {journal}
  {\bibinfo  {journal} {Phys. Rev. B}\ }\textbf {\bibinfo {volume} {99}},\
  \bibinfo {pages} {235417} (\bibinfo {year} {2019})}\BibitemShut {NoStop}%
\bibitem [{\citenamefont {{Haddadi}}\ \emph {et~al.}(2019)\citenamefont
  {{Haddadi}}, \citenamefont {{Wu}}, \citenamefont {{Kruchkov}},\ and\
  \citenamefont {{Yazyev}}}]{arxiv.Oleg}%
  \BibitemOpen
  \bibfield  {author} {\bibinfo {author} {\bibfnamefont {F.}~\bibnamefont
  {{Haddadi}}}, \bibinfo {author} {\bibfnamefont {Q.}~\bibnamefont {{Wu}}},
  \bibinfo {author} {\bibfnamefont {A.~J.}\ \bibnamefont {{Kruchkov}}}, \ and\
  \bibinfo {author} {\bibfnamefont {O.~V.}\ \bibnamefont {{Yazyev}}},\
  }\href@noop {} {\bibfield  {journal} {\bibinfo  {journal} {arXiv e-prints}\
  ,\ \bibinfo {eid} {arXiv:1906.00623}} (\bibinfo {year} {2019})},\ \Eprint
  {http://arxiv.org/abs/1906.00623} {arXiv:1906.00623 [cond-mat.mes-hall]}
  \BibitemShut {NoStop}%
\bibitem [{\citenamefont {Xian}\ \emph {et~al.}(0)\citenamefont {Xian},
  \citenamefont {Kennes}, \citenamefont {Tancogne-Dejean}, \citenamefont
  {Altarelli},\ and\ \citenamefont {Rubio}}]{NL.Xian}%
  \BibitemOpen
  \bibfield  {author} {\bibinfo {author} {\bibfnamefont {L.}~\bibnamefont
  {Xian}}, \bibinfo {author} {\bibfnamefont {D.~M.}\ \bibnamefont {Kennes}},
  \bibinfo {author} {\bibfnamefont {N.}~\bibnamefont {Tancogne-Dejean}},
  \bibinfo {author} {\bibfnamefont {M.}~\bibnamefont {Altarelli}}, \ and\
  \bibinfo {author} {\bibfnamefont {A.}~\bibnamefont {Rubio}},\ }\href
  {\doibase 10.1021/acs.nanolett.9b00986} {\bibfield  {journal} {\bibinfo
  {journal} {Nano Letters}\ }\textbf {\bibinfo {volume} {0}},\ \bibinfo {pages}
  {null} (\bibinfo {year} {0})}\BibitemShut {NoStop}%
\bibitem [{\citenamefont {Kang}\ \emph {et~al.}(2017)\citenamefont {Kang},
  \citenamefont {Zhang}, \citenamefont {Michaud-Rioux}, \citenamefont {Kong},
  \citenamefont {Hu}, \citenamefont {Yu},\ and\ \citenamefont
  {Guo}}]{PRB.Kang}%
  \BibitemOpen
  \bibfield  {author} {\bibinfo {author} {\bibfnamefont {P.}~\bibnamefont
  {Kang}}, \bibinfo {author} {\bibfnamefont {W.-T.}\ \bibnamefont {Zhang}},
  \bibinfo {author} {\bibfnamefont {V.}~\bibnamefont {Michaud-Rioux}}, \bibinfo
  {author} {\bibfnamefont {X.-H.}\ \bibnamefont {Kong}}, \bibinfo {author}
  {\bibfnamefont {C.}~\bibnamefont {Hu}}, \bibinfo {author} {\bibfnamefont
  {G.-H.}\ \bibnamefont {Yu}}, \ and\ \bibinfo {author} {\bibfnamefont
  {H.}~\bibnamefont {Guo}},\ }\href@noop {} {\bibfield  {journal} {\bibinfo
  {journal} {Phys. Rev. B}\ }\textbf {\bibinfo {volume} {96}},\ \bibinfo
  {pages} {195406} (\bibinfo {year} {2017})}\BibitemShut {NoStop}%
\bibitem [{\citenamefont {{Kennes}}\ \emph {et~al.}(2019)\citenamefont
  {{Kennes}}, \citenamefont {{Xian}}, \citenamefont {{Claassen}},\ and\
  \citenamefont {{Rubio}}}]{arxiv.Kennes}%
  \BibitemOpen
  \bibfield  {author} {\bibinfo {author} {\bibfnamefont {D.~M.}\ \bibnamefont
  {{Kennes}}}, \bibinfo {author} {\bibfnamefont {L.}~\bibnamefont {{Xian}}},
  \bibinfo {author} {\bibfnamefont {M.}~\bibnamefont {{Claassen}}}, \ and\
  \bibinfo {author} {\bibfnamefont {A.}~\bibnamefont {{Rubio}}},\ }\href@noop
  {} {\bibfield  {journal} {\bibinfo  {journal} {arXiv e-prints}\ ,\ \bibinfo
  {eid} {arXiv:1905.04025}} (\bibinfo {year} {2019})},\ \Eprint
  {http://arxiv.org/abs/1905.04025} {1905.04025} \BibitemShut {NoStop}%
\bibitem [{\citenamefont {{Zhao}}\ \emph {et~al.}(2019)\citenamefont {{Zhao}},
  \citenamefont {{Yang}}, \citenamefont {{Zhang}},\ and\ \citenamefont
  {{Wei}}}]{arxiv.Zhao}%
  \BibitemOpen
  \bibfield  {author} {\bibinfo {author} {\bibfnamefont {X.-J.}\ \bibnamefont
  {{Zhao}}}, \bibinfo {author} {\bibfnamefont {Y.}~\bibnamefont {{Yang}}},
  \bibinfo {author} {\bibfnamefont {D.-B.}\ \bibnamefont {{Zhang}}}, \ and\
  \bibinfo {author} {\bibfnamefont {S.-H.}\ \bibnamefont {{Wei}}},\ }\href@noop
  {} {\bibfield  {journal} {\bibinfo  {journal} {arXiv e-prints}\ ,\ \bibinfo
  {eid} {arXiv:1906.05992}} (\bibinfo {year} {2019})},\ \Eprint
  {http://arxiv.org/abs/1906.05992} {arXiv:1906.05992} \BibitemShut {NoStop}%
\bibitem [{\citenamefont {Naik}\ and\ \citenamefont {Jain}(2018)}]{PRL.Naik}%
  \BibitemOpen
  \bibfield  {author} {\bibinfo {author} {\bibfnamefont {M.~H.}\ \bibnamefont
  {Naik}}\ and\ \bibinfo {author} {\bibfnamefont {M.}~\bibnamefont {Jain}},\
  }\href {\doibase 10.1103/PhysRevLett.121.266401} {\bibfield  {journal}
  {\bibinfo  {journal} {Phys. Rev. Lett.}\ }\textbf {\bibinfo {volume} {121}},\
  \bibinfo {pages} {266401} (\bibinfo {year} {2018})}\BibitemShut {NoStop}%
\bibitem [{\citenamefont {Wu}\ \emph {et~al.}(2018{\natexlab{b}})\citenamefont
  {Wu}, \citenamefont {Lovorn}, \citenamefont {Tutuc},\ and\ \citenamefont
  {MacDonald}}]{PRL.MacD}%
  \BibitemOpen
  \bibfield  {author} {\bibinfo {author} {\bibfnamefont {F.}~\bibnamefont
  {Wu}}, \bibinfo {author} {\bibfnamefont {T.}~\bibnamefont {Lovorn}}, \bibinfo
  {author} {\bibfnamefont {E.}~\bibnamefont {Tutuc}}, \ and\ \bibinfo {author}
  {\bibfnamefont {A.~H.}\ \bibnamefont {MacDonald}},\ }\href {\doibase
  10.1103/PhysRevLett.121.026402} {\bibfield  {journal} {\bibinfo  {journal}
  {Phys. Rev. Lett.}\ }\textbf {\bibinfo {volume} {121}},\ \bibinfo {pages}
  {026402} (\bibinfo {year} {2018}{\natexlab{b}})}\BibitemShut {NoStop}%
\bibitem [{\citenamefont {Carr}\ \emph {et~al.}(2019)\citenamefont {Carr},
  \citenamefont {Fang}, \citenamefont {Zhu},\ and\ \citenamefont
  {Kaxiras}}]{PRR.Carr}%
  \BibitemOpen
  \bibfield  {author} {\bibinfo {author} {\bibfnamefont {S.}~\bibnamefont
  {Carr}}, \bibinfo {author} {\bibfnamefont {S.}~\bibnamefont {Fang}}, \bibinfo
  {author} {\bibfnamefont {Z.}~\bibnamefont {Zhu}}, \ and\ \bibinfo {author}
  {\bibfnamefont {E.}~\bibnamefont {Kaxiras}},\ }\href {\doibase
  10.1103/PhysRevResearch.1.013001} {\bibfield  {journal} {\bibinfo  {journal}
  {Phys. Rev. Research}\ }\textbf {\bibinfo {volume} {1}},\ \bibinfo {pages}
  {013001} (\bibinfo {year} {2019})}\BibitemShut {NoStop}%
\bibitem [{\citenamefont {{Goodwin}}\ \emph {et~al.}(2019)\citenamefont
  {{Goodwin}}, \citenamefont {{Corsetti}}, \citenamefont {{Mostofi}},\ and\
  \citenamefont {{Lischner}}}]{arxiv.Goodwin}%
  \BibitemOpen
  \bibfield  {author} {\bibinfo {author} {\bibfnamefont {Z.~A.~H.}\
  \bibnamefont {{Goodwin}}}, \bibinfo {author} {\bibfnamefont {F.}~\bibnamefont
  {{Corsetti}}}, \bibinfo {author} {\bibfnamefont {A.~A.}\ \bibnamefont
  {{Mostofi}}}, \ and\ \bibinfo {author} {\bibfnamefont {J.}~\bibnamefont
  {{Lischner}}},\ }\href@noop {} {\bibfield  {journal} {\bibinfo  {journal}
  {arXiv e-prints}\ ,\ \bibinfo {eid} {arXiv:1905.01887}} (\bibinfo {year}
  {2019})},\ \Eprint {http://arxiv.org/abs/1905.01887} {arXiv:1905.01887}
  \BibitemShut {NoStop}%
\bibitem [{\citenamefont {Tran}\ \emph
  {et~al.}(2019{\natexlab{a}})\citenamefont {Tran}, \citenamefont {Moody},
  \citenamefont {Wu}, \citenamefont {Lu}, \citenamefont {Choi}, \citenamefont
  {Kim}, \citenamefont {Rai}, \citenamefont {Sanchez}, \citenamefont {Quan},
  \citenamefont {Singh}, \citenamefont {Embley}, \citenamefont {Zepeda},
  \citenamefont {Campbell}, \citenamefont {Autry}, \citenamefont {Taniguchi},
  \citenamefont {Watanabe}, \citenamefont {Lu}, \citenamefont {Banerjee},
  \citenamefont {Silverman}, \citenamefont {Kim}, \citenamefont {Tutuc},
  \citenamefont {Yang}, \citenamefont {MacDonald},\ and\ \citenamefont
  {Li}}]{Nature.Tran}%
  \BibitemOpen
  \bibfield  {author} {\bibinfo {author} {\bibfnamefont {K.}~\bibnamefont
  {Tran}}, \bibinfo {author} {\bibfnamefont {G.}~\bibnamefont {Moody}},
  \bibinfo {author} {\bibfnamefont {F.}~\bibnamefont {Wu}}, \bibinfo {author}
  {\bibfnamefont {X.}~\bibnamefont {Lu}}, \bibinfo {author} {\bibfnamefont
  {J.}~\bibnamefont {Choi}}, \bibinfo {author} {\bibfnamefont {K.}~\bibnamefont
  {Kim}}, \bibinfo {author} {\bibfnamefont {A.}~\bibnamefont {Rai}}, \bibinfo
  {author} {\bibfnamefont {D.~A.}\ \bibnamefont {Sanchez}}, \bibinfo {author}
  {\bibfnamefont {J.}~\bibnamefont {Quan}}, \bibinfo {author} {\bibfnamefont
  {A.}~\bibnamefont {Singh}}, \bibinfo {author} {\bibfnamefont
  {J.}~\bibnamefont {Embley}}, \bibinfo {author} {\bibfnamefont
  {A.}~\bibnamefont {Zepeda}}, \bibinfo {author} {\bibfnamefont
  {M.}~\bibnamefont {Campbell}}, \bibinfo {author} {\bibfnamefont
  {T.}~\bibnamefont {Autry}}, \bibinfo {author} {\bibfnamefont
  {T.}~\bibnamefont {Taniguchi}}, \bibinfo {author} {\bibfnamefont
  {K.}~\bibnamefont {Watanabe}}, \bibinfo {author} {\bibfnamefont
  {N.}~\bibnamefont {Lu}}, \bibinfo {author} {\bibfnamefont {S.~K.}\
  \bibnamefont {Banerjee}}, \bibinfo {author} {\bibfnamefont {K.~L.}\
  \bibnamefont {Silverman}}, \bibinfo {author} {\bibfnamefont {S.}~\bibnamefont
  {Kim}}, \bibinfo {author} {\bibfnamefont {E.}~\bibnamefont {Tutuc}}, \bibinfo
  {author} {\bibfnamefont {L.}~\bibnamefont {Yang}}, \bibinfo {author}
  {\bibfnamefont {A.~H.}\ \bibnamefont {MacDonald}}, \ and\ \bibinfo {author}
  {\bibfnamefont {X.}~\bibnamefont {Li}},\ }\href@noop {} {\bibfield  {journal}
  {\bibinfo  {journal} {Nature}\ }\textbf {\bibinfo {volume} {567}},\ \bibinfo
  {pages} {71} (\bibinfo {year} {2019}{\natexlab{a}})}\BibitemShut {NoStop}%
\bibitem [{\citenamefont {Alexeev}\ \emph {et~al.}(2019)\citenamefont
  {Alexeev}, \citenamefont {Ruiz-Tijerina}, \citenamefont {Danovich},
  \citenamefont {Hamer}, \citenamefont {Terry}, \citenamefont {Nayak},
  \citenamefont {Ahn}, \citenamefont {Pak}, \citenamefont {Lee}, \citenamefont
  {Sohn}, \citenamefont {Molas}, \citenamefont {Koperski}, \citenamefont
  {Watanabe}, \citenamefont {Taniguchi}, \citenamefont {Novoselov},
  \citenamefont {Gorbachev}, \citenamefont {Shin}, \citenamefont {Fal'ko},\
  and\ \citenamefont {Tartakovskii}}]{Nature.Alexeev}%
  \BibitemOpen
  \bibfield  {author} {\bibinfo {author} {\bibfnamefont {E.~M.}\ \bibnamefont
  {Alexeev}}, \bibinfo {author} {\bibfnamefont {D.~A.}\ \bibnamefont
  {Ruiz-Tijerina}}, \bibinfo {author} {\bibfnamefont {M.}~\bibnamefont
  {Danovich}}, \bibinfo {author} {\bibfnamefont {M.~J.}\ \bibnamefont {Hamer}},
  \bibinfo {author} {\bibfnamefont {D.~J.}\ \bibnamefont {Terry}}, \bibinfo
  {author} {\bibfnamefont {P.~K.}\ \bibnamefont {Nayak}}, \bibinfo {author}
  {\bibfnamefont {S.}~\bibnamefont {Ahn}}, \bibinfo {author} {\bibfnamefont
  {S.}~\bibnamefont {Pak}}, \bibinfo {author} {\bibfnamefont {J.}~\bibnamefont
  {Lee}}, \bibinfo {author} {\bibfnamefont {J.~I.}\ \bibnamefont {Sohn}},
  \bibinfo {author} {\bibfnamefont {M.~R.}\ \bibnamefont {Molas}}, \bibinfo
  {author} {\bibfnamefont {M.}~\bibnamefont {Koperski}}, \bibinfo {author}
  {\bibfnamefont {K.}~\bibnamefont {Watanabe}}, \bibinfo {author}
  {\bibfnamefont {T.}~\bibnamefont {Taniguchi}}, \bibinfo {author}
  {\bibfnamefont {K.~S.}\ \bibnamefont {Novoselov}}, \bibinfo {author}
  {\bibfnamefont {R.~V.}\ \bibnamefont {Gorbachev}}, \bibinfo {author}
  {\bibfnamefont {H.~S.}\ \bibnamefont {Shin}}, \bibinfo {author}
  {\bibfnamefont {V.~I.}\ \bibnamefont {Fal'ko}}, \ and\ \bibinfo {author}
  {\bibfnamefont {A.~I.}\ \bibnamefont {Tartakovskii}},\ }\href@noop {}
  {\bibfield  {journal} {\bibinfo  {journal} {Nature}\ }\textbf {\bibinfo
  {volume} {567}},\ \bibinfo {pages} {81} (\bibinfo {year} {2019})}\BibitemShut
  {NoStop}%
\bibitem [{\citenamefont {Gargiulo}\ and\ \citenamefont
  {Yazyev}(2018)}]{2D.Oleg}%
  \BibitemOpen
  \bibfield  {author} {\bibinfo {author} {\bibfnamefont {F.}~\bibnamefont
  {Gargiulo}}\ and\ \bibinfo {author} {\bibfnamefont {O.~V.}\ \bibnamefont
  {Yazyev}},\ }\href@noop {} {\bibfield  {journal} {\bibinfo  {journal} {2D
  Materials}\ }\textbf {\bibinfo {volume} {5}},\ \bibinfo {pages} {015019}
  (\bibinfo {year} {2018})}\BibitemShut {NoStop}%
\bibitem [{\citenamefont {Naik}\ \emph {et~al.}(2019)\citenamefont {Naik},
  \citenamefont {Maity}, \citenamefont {Maiti},\ and\ \citenamefont
  {Jain}}]{KC.Naik}%
  \BibitemOpen
  \bibfield  {author} {\bibinfo {author} {\bibfnamefont {M.~H.}\ \bibnamefont
  {Naik}}, \bibinfo {author} {\bibfnamefont {I.}~\bibnamefont {Maity}},
  \bibinfo {author} {\bibfnamefont {P.~K.}\ \bibnamefont {Maiti}}, \ and\
  \bibinfo {author} {\bibfnamefont {M.}~\bibnamefont {Jain}},\ }\href {\doibase
  10.1021/acs.jpcc.8b10392} {\bibfield  {journal} {\bibinfo  {journal} {The
  Journal of Physical Chemistry C}\ }\textbf {\bibinfo {volume} {123}},\
  \bibinfo {pages} {9770} (\bibinfo {year} {2019})}\BibitemShut {NoStop}%
\bibitem [{\citenamefont {Maity}\ \emph {et~al.}(2020)\citenamefont {Maity},
  \citenamefont {Naik}, \citenamefont {Maiti}, \citenamefont {Krishnamurthy},\
  and\ \citenamefont {Jain}}]{arxiv.Maity}%
  \BibitemOpen
  \bibfield  {author} {\bibinfo {author} {\bibfnamefont {I.}~\bibnamefont
  {Maity}}, \bibinfo {author} {\bibfnamefont {M.~H.}\ \bibnamefont {Naik}},
  \bibinfo {author} {\bibfnamefont {P.~K.}\ \bibnamefont {Maiti}}, \bibinfo
  {author} {\bibfnamefont {H.~R.}\ \bibnamefont {Krishnamurthy}}, \ and\
  \bibinfo {author} {\bibfnamefont {M.}~\bibnamefont {Jain}},\ }\href {\doibase
  10.1103/PhysRevResearch.2.013335} {\bibfield  {journal} {\bibinfo  {journal}
  {Phys. Rev. Research}\ }\textbf {\bibinfo {volume} {2}},\ \bibinfo {pages}
  {013335} (\bibinfo {year} {2020})}\BibitemShut {NoStop}%
\bibitem [{\citenamefont {Carr}\ \emph {et~al.}(2018)\citenamefont {Carr},
  \citenamefont {Massatt}, \citenamefont {Torrisi}, \citenamefont {Cazeaux},
  \citenamefont {Luskin},\ and\ \citenamefont {Kaxiras}}]{PRB.Carr_Soliton}%
  \BibitemOpen
  \bibfield  {author} {\bibinfo {author} {\bibfnamefont {S.}~\bibnamefont
  {Carr}}, \bibinfo {author} {\bibfnamefont {D.}~\bibnamefont {Massatt}},
  \bibinfo {author} {\bibfnamefont {S.~B.}\ \bibnamefont {Torrisi}}, \bibinfo
  {author} {\bibfnamefont {P.}~\bibnamefont {Cazeaux}}, \bibinfo {author}
  {\bibfnamefont {M.}~\bibnamefont {Luskin}}, \ and\ \bibinfo {author}
  {\bibfnamefont {E.}~\bibnamefont {Kaxiras}},\ }\href {\doibase
  10.1103/PhysRevB.98.224102} {\bibfield  {journal} {\bibinfo  {journal} {Phys.
  Rev. B}\ }\textbf {\bibinfo {volume} {98}},\ \bibinfo {pages} {224102}
  (\bibinfo {year} {2018})}\BibitemShut {NoStop}%
\bibitem [{\citenamefont {Weston}\ \emph {et~al.}(2020)\citenamefont {Weston},
  \citenamefont {Zou}, \citenamefont {Enaldiev}, \citenamefont {Summerfield},
  \citenamefont {Clark}, \citenamefont {Z{\'o}lyomi}, \citenamefont {Graham},
  \citenamefont {Yelgel}, \citenamefont {Magorrian}, \citenamefont {Zhou},
  \citenamefont {Zultak}, \citenamefont {Hopkinson}, \citenamefont {Barinov},
  \citenamefont {Bointon}, \citenamefont {Kretinin}, \citenamefont {Wilson},
  \citenamefont {Beton}, \citenamefont {Fal'ko}, \citenamefont {Haigh},\ and\
  \citenamefont {Gorbachev}}]{NN.Weston}%
  \BibitemOpen
  \bibfield  {author} {\bibinfo {author} {\bibfnamefont {A.}~\bibnamefont
  {Weston}}, \bibinfo {author} {\bibfnamefont {Y.}~\bibnamefont {Zou}},
  \bibinfo {author} {\bibfnamefont {V.}~\bibnamefont {Enaldiev}}, \bibinfo
  {author} {\bibfnamefont {A.}~\bibnamefont {Summerfield}}, \bibinfo {author}
  {\bibfnamefont {N.}~\bibnamefont {Clark}}, \bibinfo {author} {\bibfnamefont
  {V.}~\bibnamefont {Z{\'o}lyomi}}, \bibinfo {author} {\bibfnamefont
  {A.}~\bibnamefont {Graham}}, \bibinfo {author} {\bibfnamefont
  {C.}~\bibnamefont {Yelgel}}, \bibinfo {author} {\bibfnamefont
  {S.}~\bibnamefont {Magorrian}}, \bibinfo {author} {\bibfnamefont
  {M.}~\bibnamefont {Zhou}}, \bibinfo {author} {\bibfnamefont {J.}~\bibnamefont
  {Zultak}}, \bibinfo {author} {\bibfnamefont {D.}~\bibnamefont {Hopkinson}},
  \bibinfo {author} {\bibfnamefont {A.}~\bibnamefont {Barinov}}, \bibinfo
  {author} {\bibfnamefont {T.~H.}\ \bibnamefont {Bointon}}, \bibinfo {author}
  {\bibfnamefont {A.}~\bibnamefont {Kretinin}}, \bibinfo {author}
  {\bibfnamefont {N.~R.}\ \bibnamefont {Wilson}}, \bibinfo {author}
  {\bibfnamefont {P.~H.}\ \bibnamefont {Beton}}, \bibinfo {author}
  {\bibfnamefont {V.~I.}\ \bibnamefont {Fal'ko}}, \bibinfo {author}
  {\bibfnamefont {S.~J.}\ \bibnamefont {Haigh}}, \ and\ \bibinfo {author}
  {\bibfnamefont {R.}~\bibnamefont {Gorbachev}},\ }\href@noop {} {\bibfield
  {journal} {\bibinfo  {journal} {Nature Nanotechnology}\ } (\bibinfo {year}
  {2020})}\BibitemShut {NoStop}%
\bibitem [{\citenamefont {Rosenberger}\ \emph {et~al.}(2020)\citenamefont
  {Rosenberger}, \citenamefont {Chuang}, \citenamefont {Phillips},
  \citenamefont {Oleshko}, \citenamefont {McCreary}, \citenamefont {Sivaram},
  \citenamefont {Hellberg},\ and\ \citenamefont {Jonker}}]{AN.Rosenberger}%
  \BibitemOpen
  \bibfield  {author} {\bibinfo {author} {\bibfnamefont {M.~R.}\ \bibnamefont
  {Rosenberger}}, \bibinfo {author} {\bibfnamefont {H.-J.}\ \bibnamefont
  {Chuang}}, \bibinfo {author} {\bibfnamefont {M.}~\bibnamefont {Phillips}},
  \bibinfo {author} {\bibfnamefont {V.~P.}\ \bibnamefont {Oleshko}}, \bibinfo
  {author} {\bibfnamefont {K.~M.}\ \bibnamefont {McCreary}}, \bibinfo {author}
  {\bibfnamefont {S.~V.}\ \bibnamefont {Sivaram}}, \bibinfo {author}
  {\bibfnamefont {C.~S.}\ \bibnamefont {Hellberg}}, \ and\ \bibinfo {author}
  {\bibfnamefont {B.~T.}\ \bibnamefont {Jonker}},\ }\href {\doibase
  10.1021/acsnano.0c00088} {\bibfield  {journal} {\bibinfo  {journal} {ACS
  Nano}\ }\textbf {\bibinfo {volume} {14}},\ \bibinfo {pages} {4550} (\bibinfo
  {year} {2020})},\ \bibinfo {note} {pMID: 32167748}\BibitemShut {NoStop}%
\bibitem [{\citenamefont {Ashoori}(1996)}]{Nat.Ashoori}%
  \BibitemOpen
  \bibfield  {author} {\bibinfo {author} {\bibfnamefont {R.~C.}\ \bibnamefont
  {Ashoori}},\ }\href@noop {} {\bibfield  {journal} {\bibinfo  {journal}
  {Nature}\ }\textbf {\bibinfo {volume} {379}},\ \bibinfo {pages} {413}
  (\bibinfo {year} {1996})}\BibitemShut {NoStop}%
\bibitem [{\citenamefont {Banin}\ \emph {et~al.}(1999)\citenamefont {Banin},
  \citenamefont {Cao}, \citenamefont {Katz},\ and\ \citenamefont
  {Millo}}]{Nat.Banin}%
  \BibitemOpen
  \bibfield  {author} {\bibinfo {author} {\bibfnamefont {U.}~\bibnamefont
  {Banin}}, \bibinfo {author} {\bibfnamefont {Y.}~\bibnamefont {Cao}}, \bibinfo
  {author} {\bibfnamefont {D.}~\bibnamefont {Katz}}, \ and\ \bibinfo {author}
  {\bibfnamefont {O.}~\bibnamefont {Millo}},\ }\href@noop {} {\bibfield
  {journal} {\bibinfo  {journal} {Nature}\ }\textbf {\bibinfo {volume} {400}},\
  \bibinfo {pages} {542} (\bibinfo {year} {1999})}\BibitemShut {NoStop}%
\bibitem [{\citenamefont {Yu}\ \emph {et~al.}(2017)\citenamefont {Yu},
  \citenamefont {Liu}, \citenamefont {Tang}, \citenamefont {Xu},\ and\
  \citenamefont {Yao}}]{Science.Yu}%
  \BibitemOpen
  \bibfield  {author} {\bibinfo {author} {\bibfnamefont {H.}~\bibnamefont
  {Yu}}, \bibinfo {author} {\bibfnamefont {G.-B.}\ \bibnamefont {Liu}},
  \bibinfo {author} {\bibfnamefont {J.}~\bibnamefont {Tang}}, \bibinfo {author}
  {\bibfnamefont {X.}~\bibnamefont {Xu}}, \ and\ \bibinfo {author}
  {\bibfnamefont {W.}~\bibnamefont {Yao}},\ }\href {\doibase
  10.1126/sciadv.1701696} {\bibfield  {journal} {\bibinfo  {journal} {Science
  Advances}\ }\textbf {\bibinfo {volume} {3}} (\bibinfo {year} {2017}),\
  10.1126/sciadv.1701696}\BibitemShut {NoStop}%
\bibitem [{\citenamefont {Xu}\ \emph {et~al.}(2015)\citenamefont {Xu},
  \citenamefont {Li},\ and\ \citenamefont {Wu}}]{AFM.Xu}%
  \BibitemOpen
  \bibfield  {author} {\bibinfo {author} {\bibfnamefont {S.}~\bibnamefont
  {Xu}}, \bibinfo {author} {\bibfnamefont {D.}~\bibnamefont {Li}}, \ and\
  \bibinfo {author} {\bibfnamefont {P.}~\bibnamefont {Wu}},\ }\href {\doibase
  10.1002/adfm.201403863} {\bibfield  {journal} {\bibinfo  {journal} {Advanced
  Functional Materials}\ }\textbf {\bibinfo {volume} {25}},\ \bibinfo {pages}
  {1127} (\bibinfo {year} {2015})}\BibitemShut {NoStop}%
\bibitem [{\citenamefont {Gopalakrishnan}\ \emph {et~al.}(2014)\citenamefont
  {Gopalakrishnan}, \citenamefont {Damien},\ and\ \citenamefont
  {Shaijumon}}]{AN.Deepesh}%
  \BibitemOpen
  \bibfield  {author} {\bibinfo {author} {\bibfnamefont {D.}~\bibnamefont
  {Gopalakrishnan}}, \bibinfo {author} {\bibfnamefont {D.}~\bibnamefont
  {Damien}}, \ and\ \bibinfo {author} {\bibfnamefont {M.~M.}\ \bibnamefont
  {Shaijumon}},\ }\href {\doibase 10.1021/nn501479e} {\bibfield  {journal}
  {\bibinfo  {journal} {ACS Nano}\ }\textbf {\bibinfo {volume} {8}},\ \bibinfo
  {pages} {5297} (\bibinfo {year} {2014})}\BibitemShut {NoStop}%
\bibitem [{\citenamefont {Gan}\ \emph {et~al.}(2016)\citenamefont {Gan},
  \citenamefont {Gui}, \citenamefont {Shan}, \citenamefont {Pan}, \citenamefont
  {Zhang},\ and\ \citenamefont {Zhang}}]{JAP.Gan}%
  \BibitemOpen
  \bibfield  {author} {\bibinfo {author} {\bibfnamefont {Z.}~\bibnamefont
  {Gan}}, \bibinfo {author} {\bibfnamefont {Q.}~\bibnamefont {Gui}}, \bibinfo
  {author} {\bibfnamefont {Y.}~\bibnamefont {Shan}}, \bibinfo {author}
  {\bibfnamefont {P.}~\bibnamefont {Pan}}, \bibinfo {author} {\bibfnamefont
  {N.}~\bibnamefont {Zhang}}, \ and\ \bibinfo {author} {\bibfnamefont
  {L.}~\bibnamefont {Zhang}},\ }\href {\doibase 10.1063/1.4962318} {\bibfield
  {journal} {\bibinfo  {journal} {Journal of Applied Physics}\ }\textbf
  {\bibinfo {volume} {120}},\ \bibinfo {pages} {104503} (\bibinfo {year}
  {2016})}\BibitemShut {NoStop}%
\bibitem [{\citenamefont {Perumal~Veeramalai}\ \emph
  {et~al.}(2019)\citenamefont {Perumal~Veeramalai}, \citenamefont {Li},
  \citenamefont {Guo},\ and\ \citenamefont {Kim}}]{DT.Perumal}%
  \BibitemOpen
  \bibfield  {author} {\bibinfo {author} {\bibfnamefont {C.}~\bibnamefont
  {Perumal~Veeramalai}}, \bibinfo {author} {\bibfnamefont {F.}~\bibnamefont
  {Li}}, \bibinfo {author} {\bibfnamefont {T.}~\bibnamefont {Guo}}, \ and\
  \bibinfo {author} {\bibfnamefont {T.~W.}\ \bibnamefont {Kim}},\ }\href
  {\doibase 10.1039/C8DT04593C} {\bibfield  {journal} {\bibinfo  {journal}
  {Dalton Trans.}\ }\textbf {\bibinfo {volume} {48}},\ \bibinfo {pages} {2422}
  (\bibinfo {year} {2019})}\BibitemShut {NoStop}%
\bibitem [{\citenamefont {Ha}\ \emph {et~al.}(2014)\citenamefont {Ha},
  \citenamefont {Han}, \citenamefont {Choi}, \citenamefont {Park},\ and\
  \citenamefont {Seo}}]{Small.Ha}%
  \BibitemOpen
  \bibfield  {author} {\bibinfo {author} {\bibfnamefont {H.~D.}\ \bibnamefont
  {Ha}}, \bibinfo {author} {\bibfnamefont {D.~J.}\ \bibnamefont {Han}},
  \bibinfo {author} {\bibfnamefont {J.~S.}\ \bibnamefont {Choi}}, \bibinfo
  {author} {\bibfnamefont {M.}~\bibnamefont {Park}}, \ and\ \bibinfo {author}
  {\bibfnamefont {T.~S.}\ \bibnamefont {Seo}},\ }\href {\doibase
  10.1002/smll.201400988} {\bibfield  {journal} {\bibinfo  {journal} {Small}\
  }\textbf {\bibinfo {volume} {10}},\ \bibinfo {pages} {3858} (\bibinfo {year}
  {2014})}\BibitemShut {NoStop}%
\bibitem [{\citenamefont {Lin}\ \emph {et~al.}(2015)\citenamefont {Lin},
  \citenamefont {Wang}, \citenamefont {Wu}, \citenamefont {Xu}, \citenamefont
  {Huang},\ and\ \citenamefont {Zhang}}]{NJC.Lin}%
  \BibitemOpen
  \bibfield  {author} {\bibinfo {author} {\bibfnamefont {H.}~\bibnamefont
  {Lin}}, \bibinfo {author} {\bibfnamefont {C.}~\bibnamefont {Wang}}, \bibinfo
  {author} {\bibfnamefont {J.}~\bibnamefont {Wu}}, \bibinfo {author}
  {\bibfnamefont {Z.}~\bibnamefont {Xu}}, \bibinfo {author} {\bibfnamefont
  {Y.}~\bibnamefont {Huang}}, \ and\ \bibinfo {author} {\bibfnamefont
  {C.}~\bibnamefont {Zhang}},\ }\href@noop {} {\bibfield  {journal} {\bibinfo
  {journal} {New J. Chem.}\ }\textbf {\bibinfo {volume} {39}},\ \bibinfo
  {pages} {8492} (\bibinfo {year} {2015})}\BibitemShut {NoStop}%
\bibitem [{\citenamefont {Wu}\ \emph {et~al.}(2014)\citenamefont {Wu},
  \citenamefont {Qian},\ and\ \citenamefont {Li}}]{NL.Wu}%
  \BibitemOpen
  \bibfield  {author} {\bibinfo {author} {\bibfnamefont {M.}~\bibnamefont
  {Wu}}, \bibinfo {author} {\bibfnamefont {X.}~\bibnamefont {Qian}}, \ and\
  \bibinfo {author} {\bibfnamefont {J.}~\bibnamefont {Li}},\ }\href@noop {}
  {\bibfield  {journal} {\bibinfo  {journal} {Nano Letters}\ }\textbf {\bibinfo
  {volume} {14}},\ \bibinfo {pages} {5350} (\bibinfo {year}
  {2014})}\BibitemShut {NoStop}%
\bibitem [{\citenamefont {{Fleischmann}}\ \emph {et~al.}(2019)\citenamefont
  {{Fleischmann}}, \citenamefont {{Gupta}}, \citenamefont {{Sharma}},\ and\
  \citenamefont {{Shallcross}}}]{arxiv.Shallcross}%
  \BibitemOpen
  \bibfield  {author} {\bibinfo {author} {\bibfnamefont {M.}~\bibnamefont
  {{Fleischmann}}}, \bibinfo {author} {\bibfnamefont {R.}~\bibnamefont
  {{Gupta}}}, \bibinfo {author} {\bibfnamefont {S.}~\bibnamefont {{Sharma}}}, \
  and\ \bibinfo {author} {\bibfnamefont {S.}~\bibnamefont {{Shallcross}}},\
  }\href@noop {} {\bibfield  {journal} {\bibinfo  {journal} {arXiv e-prints}\
  ,\ \bibinfo {eid} {arXiv:1901.04679}} (\bibinfo {year} {2019})},\ \Eprint
  {http://arxiv.org/abs/1901.04679} {1901.04679} \BibitemShut {NoStop}%
\bibitem [{\citenamefont {Seyler}\ \emph {et~al.}(2019)\citenamefont {Seyler},
  \citenamefont {Rivera}, \citenamefont {Yu}, \citenamefont {Wilson},
  \citenamefont {Ray}, \citenamefont {Mandrus}, \citenamefont {Yan},
  \citenamefont {Yao},\ and\ \citenamefont {Xu}}]{Nat.Seyler}%
  \BibitemOpen
  \bibfield  {author} {\bibinfo {author} {\bibfnamefont {K.~L.}\ \bibnamefont
  {Seyler}}, \bibinfo {author} {\bibfnamefont {P.}~\bibnamefont {Rivera}},
  \bibinfo {author} {\bibfnamefont {H.}~\bibnamefont {Yu}}, \bibinfo {author}
  {\bibfnamefont {N.~P.}\ \bibnamefont {Wilson}}, \bibinfo {author}
  {\bibfnamefont {E.~L.}\ \bibnamefont {Ray}}, \bibinfo {author} {\bibfnamefont
  {D.~G.}\ \bibnamefont {Mandrus}}, \bibinfo {author} {\bibfnamefont
  {J.}~\bibnamefont {Yan}}, \bibinfo {author} {\bibfnamefont {W.}~\bibnamefont
  {Yao}}, \ and\ \bibinfo {author} {\bibfnamefont {X.}~\bibnamefont {Xu}},\
  }\href@noop {} {\bibfield  {journal} {\bibinfo  {journal} {Nature}\ }\textbf
  {\bibinfo {volume} {567}},\ \bibinfo {pages} {66} (\bibinfo {year}
  {2019})}\BibitemShut {NoStop}%
\bibitem [{\citenamefont {Jin}\ \emph {et~al.}(2019)\citenamefont {Jin},
  \citenamefont {Regan}, \citenamefont {Yan}, \citenamefont {Iqbal
  Bakti~Utama}, \citenamefont {Wang}, \citenamefont {Zhao}, \citenamefont
  {Qin}, \citenamefont {Yang}, \citenamefont {Zheng}, \citenamefont {Shi},
  \citenamefont {Watanabe}, \citenamefont {Taniguchi}, \citenamefont {Tongay},
  \citenamefont {Zettl},\ and\ \citenamefont {Wang}}]{Nat.Jin}%
  \BibitemOpen
  \bibfield  {author} {\bibinfo {author} {\bibfnamefont {C.}~\bibnamefont
  {Jin}}, \bibinfo {author} {\bibfnamefont {E.~C.}\ \bibnamefont {Regan}},
  \bibinfo {author} {\bibfnamefont {A.}~\bibnamefont {Yan}}, \bibinfo {author}
  {\bibfnamefont {M.}~\bibnamefont {Iqbal Bakti~Utama}}, \bibinfo {author}
  {\bibfnamefont {D.}~\bibnamefont {Wang}}, \bibinfo {author} {\bibfnamefont
  {S.}~\bibnamefont {Zhao}}, \bibinfo {author} {\bibfnamefont {Y.}~\bibnamefont
  {Qin}}, \bibinfo {author} {\bibfnamefont {S.}~\bibnamefont {Yang}}, \bibinfo
  {author} {\bibfnamefont {Z.}~\bibnamefont {Zheng}}, \bibinfo {author}
  {\bibfnamefont {S.}~\bibnamefont {Shi}}, \bibinfo {author} {\bibfnamefont
  {K.}~\bibnamefont {Watanabe}}, \bibinfo {author} {\bibfnamefont
  {T.}~\bibnamefont {Taniguchi}}, \bibinfo {author} {\bibfnamefont
  {S.}~\bibnamefont {Tongay}}, \bibinfo {author} {\bibfnamefont
  {A.}~\bibnamefont {Zettl}}, \ and\ \bibinfo {author} {\bibfnamefont
  {F.}~\bibnamefont {Wang}},\ }\href@noop {} {\bibfield  {journal} {\bibinfo
  {journal} {Nature}\ }\textbf {\bibinfo {volume} {567}},\ \bibinfo {pages}
  {76} (\bibinfo {year} {2019})}\BibitemShut {NoStop}%
\bibitem [{\citenamefont {Tran}\ \emph
  {et~al.}(2019{\natexlab{b}})\citenamefont {Tran}, \citenamefont {Moody},
  \citenamefont {Wu}, \citenamefont {Lu}, \citenamefont {Choi}, \citenamefont
  {Kim}, \citenamefont {Rai}, \citenamefont {Sanchez}, \citenamefont {Quan},
  \citenamefont {Singh}, \citenamefont {Embley}, \citenamefont {Zepeda},
  \citenamefont {Campbell}, \citenamefont {Autry}, \citenamefont {Taniguchi},
  \citenamefont {Watanabe}, \citenamefont {Lu}, \citenamefont {Banerjee},
  \citenamefont {Silverman}, \citenamefont {Kim}, \citenamefont {Tutuc},
  \citenamefont {Yang}, \citenamefont {MacDonald},\ and\ \citenamefont
  {Li}}]{Nat.Tran}%
  \BibitemOpen
  \bibfield  {author} {\bibinfo {author} {\bibfnamefont {K.}~\bibnamefont
  {Tran}}, \bibinfo {author} {\bibfnamefont {G.}~\bibnamefont {Moody}},
  \bibinfo {author} {\bibfnamefont {F.}~\bibnamefont {Wu}}, \bibinfo {author}
  {\bibfnamefont {X.}~\bibnamefont {Lu}}, \bibinfo {author} {\bibfnamefont
  {J.}~\bibnamefont {Choi}}, \bibinfo {author} {\bibfnamefont {K.}~\bibnamefont
  {Kim}}, \bibinfo {author} {\bibfnamefont {A.}~\bibnamefont {Rai}}, \bibinfo
  {author} {\bibfnamefont {D.~A.}\ \bibnamefont {Sanchez}}, \bibinfo {author}
  {\bibfnamefont {J.}~\bibnamefont {Quan}}, \bibinfo {author} {\bibfnamefont
  {A.}~\bibnamefont {Singh}}, \bibinfo {author} {\bibfnamefont
  {J.}~\bibnamefont {Embley}}, \bibinfo {author} {\bibfnamefont
  {A.}~\bibnamefont {Zepeda}}, \bibinfo {author} {\bibfnamefont
  {M.}~\bibnamefont {Campbell}}, \bibinfo {author} {\bibfnamefont
  {T.}~\bibnamefont {Autry}}, \bibinfo {author} {\bibfnamefont
  {T.}~\bibnamefont {Taniguchi}}, \bibinfo {author} {\bibfnamefont
  {K.}~\bibnamefont {Watanabe}}, \bibinfo {author} {\bibfnamefont
  {N.}~\bibnamefont {Lu}}, \bibinfo {author} {\bibfnamefont {S.~K.}\
  \bibnamefont {Banerjee}}, \bibinfo {author} {\bibfnamefont {K.~L.}\
  \bibnamefont {Silverman}}, \bibinfo {author} {\bibfnamefont {S.}~\bibnamefont
  {Kim}}, \bibinfo {author} {\bibfnamefont {E.}~\bibnamefont {Tutuc}}, \bibinfo
  {author} {\bibfnamefont {L.}~\bibnamefont {Yang}}, \bibinfo {author}
  {\bibfnamefont {A.~H.}\ \bibnamefont {MacDonald}}, \ and\ \bibinfo {author}
  {\bibfnamefont {X.}~\bibnamefont {Li}},\ }\href@noop {} {\bibfield  {journal}
  {\bibinfo  {journal} {Nature}\ }\textbf {\bibinfo {volume} {567}},\ \bibinfo
  {pages} {71} (\bibinfo {year} {2019}{\natexlab{b}})}\BibitemShut {NoStop}%
\bibitem [{\citenamefont {Regan}\ \emph {et~al.}(2019)\citenamefont {Regan},
  \citenamefont {Wang}, \citenamefont {Jin}, \citenamefont {Utama},
  \citenamefont {Gao}, \citenamefont {Wei}, \citenamefont {Zhao}, \citenamefont
  {Zhao}, \citenamefont {Yumigeta}, \citenamefont {Blei}, \citenamefont
  {Carlstroem}, \citenamefont {Watanabe}, \citenamefont {Taniguchi},
  \citenamefont {Tongay}, \citenamefont {Crommie}, \citenamefont {Zettl},\ and\
  \citenamefont {Wang}}]{arxiv.Regan}%
  \BibitemOpen
  \bibfield  {author} {\bibinfo {author} {\bibfnamefont {E.~C.}\ \bibnamefont
  {Regan}}, \bibinfo {author} {\bibfnamefont {D.}~\bibnamefont {Wang}},
  \bibinfo {author} {\bibfnamefont {C.}~\bibnamefont {Jin}}, \bibinfo {author}
  {\bibfnamefont {M.~I.~B.}\ \bibnamefont {Utama}}, \bibinfo {author}
  {\bibfnamefont {B.}~\bibnamefont {Gao}}, \bibinfo {author} {\bibfnamefont
  {X.}~\bibnamefont {Wei}}, \bibinfo {author} {\bibfnamefont {S.}~\bibnamefont
  {Zhao}}, \bibinfo {author} {\bibfnamefont {W.}~\bibnamefont {Zhao}}, \bibinfo
  {author} {\bibfnamefont {K.}~\bibnamefont {Yumigeta}}, \bibinfo {author}
  {\bibfnamefont {M.}~\bibnamefont {Blei}}, \bibinfo {author} {\bibfnamefont
  {J.}~\bibnamefont {Carlstroem}}, \bibinfo {author} {\bibfnamefont
  {K.}~\bibnamefont {Watanabe}}, \bibinfo {author} {\bibfnamefont
  {T.}~\bibnamefont {Taniguchi}}, \bibinfo {author} {\bibfnamefont
  {S.}~\bibnamefont {Tongay}}, \bibinfo {author} {\bibfnamefont
  {M.}~\bibnamefont {Crommie}}, \bibinfo {author} {\bibfnamefont
  {A.}~\bibnamefont {Zettl}}, \ and\ \bibinfo {author} {\bibfnamefont
  {F.}~\bibnamefont {Wang}},\ }\href@noop {} {\enquote {\bibinfo {title}
  {Optical detection of mott and generalized wigner crystal states in wse2/ws2
  moiré superlattices},}\ } (\bibinfo {year} {2019}),\ \Eprint
  {http://arxiv.org/abs/1910.09047} {arXiv:1910.09047 [cond-mat.mes-hall]}
  \BibitemShut {NoStop}%
\bibitem [{\citenamefont {{Brotons-Gisbert}}\ \emph {et~al.}(2019)\citenamefont
  {{Brotons-Gisbert}}, \citenamefont {{Baek}}, \citenamefont
  {{Molina-S{\'a}nchez}}, \citenamefont {{Scerri}}, \citenamefont {{White}},
  \citenamefont {{Watanabe}}, \citenamefont {{Taniguchi}}, \citenamefont
  {{Bonato}},\ and\ \citenamefont {{Gerardot}}}]{arxiv.Mauro}%
  \BibitemOpen
  \bibfield  {author} {\bibinfo {author} {\bibfnamefont {M.}~\bibnamefont
  {{Brotons-Gisbert}}}, \bibinfo {author} {\bibfnamefont {H.}~\bibnamefont
  {{Baek}}}, \bibinfo {author} {\bibfnamefont {A.}~\bibnamefont
  {{Molina-S{\'a}nchez}}}, \bibinfo {author} {\bibfnamefont {D.}~\bibnamefont
  {{Scerri}}}, \bibinfo {author} {\bibfnamefont {D.}~\bibnamefont {{White}}},
  \bibinfo {author} {\bibfnamefont {K.}~\bibnamefont {{Watanabe}}}, \bibinfo
  {author} {\bibfnamefont {T.}~\bibnamefont {{Taniguchi}}}, \bibinfo {author}
  {\bibfnamefont {C.}~\bibnamefont {{Bonato}}}, \ and\ \bibinfo {author}
  {\bibfnamefont {B.~D.}\ \bibnamefont {{Gerardot}}},\ }\href@noop {}
  {\bibfield  {journal} {\bibinfo  {journal} {arXiv e-prints}\ ,\ \bibinfo
  {eid} {arXiv:1908.03778}} (\bibinfo {year} {2019})},\ \Eprint
  {http://arxiv.org/abs/1908.03778} {1908.03778} \BibitemShut {NoStop}%
\bibitem [{\citenamefont {Naik}\ and\ \citenamefont {Jain}(2017)}]{PRB.Naik}%
  \BibitemOpen
  \bibfield  {author} {\bibinfo {author} {\bibfnamefont {M.~H.}\ \bibnamefont
  {Naik}}\ and\ \bibinfo {author} {\bibfnamefont {M.}~\bibnamefont {Jain}},\
  }\href@noop {} {\bibfield  {journal} {\bibinfo  {journal} {Phys. Rev. B}\
  }\textbf {\bibinfo {volume} {95}},\ \bibinfo {pages} {165125} (\bibinfo
  {year} {2017})}\BibitemShut {NoStop}%
\bibitem [{twi()}]{twister}%
  \BibitemOpen
  \href@noop {} {}\bibinfo {howpublished}
  {\url{http://www.physics.iisc.ernet.in/~mjain/pages/software.html}}\BibitemShut
  {NoStop}%
\bibitem [{LAM()}]{LAMMPS}%
  \BibitemOpen
  \href@noop {} {}\bibinfo {howpublished}
  {\url{https://lammps.sandia.gov}}\BibitemShut {NoStop}%
\bibitem [{\citenamefont {Plimpton}(1995)}]{JCP.Plimpton}%
  \BibitemOpen
  \bibfield  {author} {\bibinfo {author} {\bibfnamefont {S.}~\bibnamefont
  {Plimpton}},\ }\href@noop {} {\bibfield  {journal} {\bibinfo  {journal}
  {Journal of Computational Physics}\ }\textbf {\bibinfo {volume} {117}},\
  \bibinfo {pages} {1 } (\bibinfo {year} {1995})}\BibitemShut {NoStop}%
\bibitem [{\citenamefont {Stillinger}\ and\ \citenamefont
  {Weber}(1985)}]{PRB.SW}%
  \BibitemOpen
  \bibfield  {author} {\bibinfo {author} {\bibfnamefont {F.~H.}\ \bibnamefont
  {Stillinger}}\ and\ \bibinfo {author} {\bibfnamefont {T.~A.}\ \bibnamefont
  {Weber}},\ }\href@noop {} {\bibfield  {journal} {\bibinfo  {journal} {Phys.
  Rev. B}\ }\textbf {\bibinfo {volume} {31}},\ \bibinfo {pages} {5262}
  (\bibinfo {year} {1985})}\BibitemShut {NoStop}%
\bibitem [{\citenamefont {Jiang}\ and\ \citenamefont {Zhou}(2017)}]{Chap.SW}%
  \BibitemOpen
  \bibfield  {author} {\bibinfo {author} {\bibfnamefont {J.-W.}\ \bibnamefont
  {Jiang}}\ and\ \bibinfo {author} {\bibfnamefont {Y.-P.}\ \bibnamefont
  {Zhou}},\ }in\ \href@noop {} {\emph {\bibinfo {booktitle} {Handbook of
  Stillinger-Weber Potential Parameters for Two-Dimensional Atomic
  Crystals}}},\ \bibinfo {editor} {edited by\ \bibinfo {editor} {\bibfnamefont
  {J.-W.}\ \bibnamefont {Jiang}}\ and\ \bibinfo {editor} {\bibfnamefont
  {Y.-P.}\ \bibnamefont {Zhou}}}\ (\bibinfo  {publisher} {IntechOpen},\
  \bibinfo {address} {Rijeka},\ \bibinfo {year} {2017})\ Chap.~\bibinfo
  {chapter} {1}\BibitemShut {NoStop}%
\bibitem [{\citenamefont {Kohn}\ and\ \citenamefont {Sham}(1965)}]{PR.Kohn}%
  \BibitemOpen
  \bibfield  {author} {\bibinfo {author} {\bibfnamefont {W.}~\bibnamefont
  {Kohn}}\ and\ \bibinfo {author} {\bibfnamefont {L.~J.}\ \bibnamefont
  {Sham}},\ }\href@noop {} {\bibfield  {journal} {\bibinfo  {journal} {Phys.
  Rev.}\ }\textbf {\bibinfo {volume} {140}},\ \bibinfo {pages} {A1133}
  (\bibinfo {year} {1965})}\BibitemShut {NoStop}%
\bibitem [{\citenamefont {Soler}\ \emph {et~al.}(2002)\citenamefont {Soler},
  \citenamefont {Artacho}, \citenamefont {Gale}, \citenamefont {Garc{\'{\i}}a},
  \citenamefont {Junquera}, \citenamefont {Ordej{\'{o}}n},\ and\ \citenamefont
  {S{\'{a}}nchez-Portal}}]{SIESTA}%
  \BibitemOpen
  \bibfield  {author} {\bibinfo {author} {\bibfnamefont {J.~M.}\ \bibnamefont
  {Soler}}, \bibinfo {author} {\bibfnamefont {E.}~\bibnamefont {Artacho}},
  \bibinfo {author} {\bibfnamefont {J.~D.}\ \bibnamefont {Gale}}, \bibinfo
  {author} {\bibfnamefont {A.}~\bibnamefont {Garc{\'{\i}}a}}, \bibinfo {author}
  {\bibfnamefont {J.}~\bibnamefont {Junquera}}, \bibinfo {author}
  {\bibfnamefont {P.}~\bibnamefont {Ordej{\'{o}}n}}, \ and\ \bibinfo {author}
  {\bibfnamefont {D.}~\bibnamefont {S{\'{a}}nchez-Portal}},\ }\href {\doibase
  10.1088/0953-8984/14/11/302} {\bibfield  {journal} {\bibinfo  {journal}
  {Journal of Physics: Condensed Matter}\ }\textbf {\bibinfo {volume} {14}},\
  \bibinfo {pages} {2745} (\bibinfo {year} {2002})}\BibitemShut {NoStop}%
\bibitem [{\citenamefont {Troullier}\ and\ \citenamefont
  {Martins}(1991)}]{PRB.TM}%
  \BibitemOpen
  \bibfield  {author} {\bibinfo {author} {\bibfnamefont {N.}~\bibnamefont
  {Troullier}}\ and\ \bibinfo {author} {\bibfnamefont {J.~L.}\ \bibnamefont
  {Martins}},\ }\href {\doibase 10.1103/PhysRevB.43.1993} {\bibfield  {journal}
  {\bibinfo  {journal} {Phys. Rev. B}\ }\textbf {\bibinfo {volume} {43}},\
  \bibinfo {pages} {1993} (\bibinfo {year} {1991})}\BibitemShut {NoStop}%
\bibitem [{\citenamefont {van Wijk}\ \emph {et~al.}(2015)\citenamefont {van
  Wijk}, \citenamefont {Schuring}, \citenamefont {Katsnelson},\ and\
  \citenamefont {Fasolino}}]{2D.Wijk}%
  \BibitemOpen
  \bibfield  {author} {\bibinfo {author} {\bibfnamefont {M.~M.}\ \bibnamefont
  {van Wijk}}, \bibinfo {author} {\bibfnamefont {A.}~\bibnamefont {Schuring}},
  \bibinfo {author} {\bibfnamefont {M.~I.}\ \bibnamefont {Katsnelson}}, \ and\
  \bibinfo {author} {\bibfnamefont {A.}~\bibnamefont {Fasolino}},\ }\href@noop
  {} {\bibfield  {journal} {\bibinfo  {journal} {2D Materials}\ }\textbf
  {\bibinfo {volume} {2}},\ \bibinfo {pages} {034010} (\bibinfo {year}
  {2015})}\BibitemShut {NoStop}%
\bibitem [{\citenamefont {Li}\ and\ \citenamefont {Blinder}(1987)}]{JCE.Li}%
  \BibitemOpen
  \bibfield  {author} {\bibinfo {author} {\bibfnamefont {W.-K.}\ \bibnamefont
  {Li}}\ and\ \bibinfo {author} {\bibfnamefont {S.~M.}\ \bibnamefont
  {Blinder}},\ }\href {\doibase 10.1021/ed064p130} {\bibfield  {journal}
  {\bibinfo  {journal} {Journal of Chemical Education}\ }\textbf {\bibinfo
  {volume} {64}},\ \bibinfo {pages} {130} (\bibinfo {year} {1987})}\BibitemShut
  {NoStop}%
\bibitem [{\citenamefont {Krishnamurthy}\ \emph {et~al.}(1982)\citenamefont
  {Krishnamurthy}, \citenamefont {Mani},\ and\ \citenamefont
  {Verma}}]{JPA.HRK}%
  \BibitemOpen
  \bibfield  {author} {\bibinfo {author} {\bibfnamefont {H.~R.}\ \bibnamefont
  {Krishnamurthy}}, \bibinfo {author} {\bibfnamefont {H.~S.}\ \bibnamefont
  {Mani}}, \ and\ \bibinfo {author} {\bibfnamefont {H.~C.}\ \bibnamefont
  {Verma}},\ }\href {\doibase 10.1088/0305-4470/15/7/024} {\bibfield  {journal}
  {\bibinfo  {journal} {Journal of Physics A: Mathematical and General}\
  }\textbf {\bibinfo {volume} {15}},\ \bibinfo {pages} {2131} (\bibinfo {year}
  {1982})}\BibitemShut {NoStop}%
\bibitem [{\citenamefont {Zhang}\ \emph {et~al.}(2017)\citenamefont {Zhang},
  \citenamefont {Chuu}, \citenamefont {Ren}, \citenamefont {Li}, \citenamefont
  {Li}, \citenamefont {Jin}, \citenamefont {Chou},\ and\ \citenamefont
  {Shih}}]{SA.Zhang}%
  \BibitemOpen
  \bibfield  {author} {\bibinfo {author} {\bibfnamefont {C.}~\bibnamefont
  {Zhang}}, \bibinfo {author} {\bibfnamefont {C.-P.}\ \bibnamefont {Chuu}},
  \bibinfo {author} {\bibfnamefont {X.}~\bibnamefont {Ren}}, \bibinfo {author}
  {\bibfnamefont {M.-Y.}\ \bibnamefont {Li}}, \bibinfo {author} {\bibfnamefont
  {L.-J.}\ \bibnamefont {Li}}, \bibinfo {author} {\bibfnamefont
  {C.}~\bibnamefont {Jin}}, \bibinfo {author} {\bibfnamefont {M.-Y.}\
  \bibnamefont {Chou}}, \ and\ \bibinfo {author} {\bibfnamefont {C.-K.}\
  \bibnamefont {Shih}},\ }\href {\doibase 10.1126/sciadv.1601459} {\bibfield
  {journal} {\bibinfo  {journal} {Science Advances}\ }\textbf {\bibinfo
  {volume} {3}} (\bibinfo {year} {2017}),\ 10.1126/sciadv.1601459}\BibitemShut
  {NoStop}%
\bibitem [{\citenamefont {Zhang}\ \emph {et~al.}(2018)\citenamefont {Zhang},
  \citenamefont {Li}, \citenamefont {Tersoff}, \citenamefont {Han},
  \citenamefont {Su}, \citenamefont {Li}, \citenamefont {Muller},\ and\
  \citenamefont {Shih}}]{NN.Zhang}%
  \BibitemOpen
  \bibfield  {author} {\bibinfo {author} {\bibfnamefont {C.}~\bibnamefont
  {Zhang}}, \bibinfo {author} {\bibfnamefont {M.-Y.}\ \bibnamefont {Li}},
  \bibinfo {author} {\bibfnamefont {J.}~\bibnamefont {Tersoff}}, \bibinfo
  {author} {\bibfnamefont {Y.}~\bibnamefont {Han}}, \bibinfo {author}
  {\bibfnamefont {Y.}~\bibnamefont {Su}}, \bibinfo {author} {\bibfnamefont
  {L.-J.}\ \bibnamefont {Li}}, \bibinfo {author} {\bibfnamefont {D.~A.}\
  \bibnamefont {Muller}}, \ and\ \bibinfo {author} {\bibfnamefont {C.-K.}\
  \bibnamefont {Shih}},\ }\href@noop {} {\bibfield  {journal} {\bibinfo
  {journal} {Nature Nanotechnology}\ }\textbf {\bibinfo {volume} {13}},\
  \bibinfo {pages} {152} (\bibinfo {year} {2018})}\BibitemShut {NoStop}%
\bibitem [{SM_()}]{SM_ref}%
  \BibitemOpen
  \href@noop {} {}\bibinfo {note} {See Supplementary Materials for movie
  showing the evolution of the flatband localisation with
  twist-angle.}\BibitemShut {Stop}%
\end{thebibliography}

%


\end{document}